\documentclass[aps,PRApplied,twocolumn,groupedaddress]{revtex4-2}

\usepackage{amsmath}
\usepackage{commath}
\usepackage{amssymb}
\usepackage{hyperref}
\usepackage{braket}
\usepackage{graphicx}
\usepackage{breqn}

\usepackage{subcaption}
\usepackage{booktabs}
\usepackage{tabularx}
\usepackage{url}

\usepackage[normalem]{ulem}

\usepackage[dvipsnames]{xcolor}

\begin{document}


\title{Phonon-Limited Transport in 2D Materials: A Unified Approach for \textit{ab initio} Mobility and Current Calculations}


\author{Jonathan Backman}
\email[]{jbackman@iis.ee.ethz.ch}
\author{Youseung Lee}
\author{Mathieu Luisier}
\affiliation{Integrated Systems Laboratory\, ETH Z\"urich\, 8092 Z\"urich Switzerland}




\begin{abstract}
This paper presents an \textit{ab initio} methodology to account for electron-phonon interactions in 2D materials, focusing on transition metal dichalcogenides (TMDCs). It combines density functional theory and maximally localized Wannier functions to acquire material data and relies on the linearized Boltzmann transport equation (LBTE) and the non-equilibrium Green’s functions (NEGF) method to determine the transport properties of materials and devices, respectively. It is shown that for MoS$_2$, both LBTE and NEGF return very close mobility values, without the need to adjust any parameter. The excellent agreement between both approaches results from the inclusion of non-diagonal entries in the electron-phonon scattering self-energies. The NEGF solver is then used to shed light on the "current vs. voltage" characteristics of a monolayer MoS$_2$ transistor, highlighting how the interactions with phonons impact both the current magnitude and its distribution. The mobility of other TMDCs is considered as well, demonstrating the capabilities of the proposed technique to assess the potential of 2D channel materials in next-generation logic applications.
\end{abstract}

\maketitle

\section{Introduction}\label{sec:intro}

More than 60 years of electronic device miniaturization have pushed Silicon, the semiconductor of reference, towards its intrinsic limit, thus calling for the emergence of alternative channel materials. Two-dimensional (2D) compounds such as transition metal dichalcogenides (TMDCs) are seen as promising candidates to equip future electronic products thanks to their unique electronic \cite{Wang2012}, thermal \cite{Kim2021}, and mechanical \cite{Akinwande2017} properties. The first demonstration of a single-layer molybdenum disulfide (MoS$_2$) transistor \cite{Radisavljevic2011} moved the attention of the device community towards single- and few-layer TMDCs as channel of logic switches. Since then, motivated by technological advancements in the growth and exfoliation of 2D semiconductors \cite{Lee2010,Kang2015}, transistors based on a wide range of TMDCs such as WSe2 \cite{Fang2012}, WS2 \cite{Ovchinnikov2014}, MoTe2 \cite{Fathipour2014}, MoSe2 \cite{Meng2017}, ReS2 \cite{Liu2015ReS2}, HfSe2, or ZrSe2 \cite{Mleczko2017} have been reported. Notably, a recent work demonstrated a nano-sheet field-effect transistor (FET) in a gate-all-around configuration with a monolayer MoS$_2$ channel \cite{Chung2022}. Despite the progress that has been made in the fabrication of 2D FETs, these components remain far from reaching their ultimate potential in terms of carrier transport. The limits of the latter properties are very often assessed through theoretical investigations, which allows for the elimination of non-intrinsic factors, e.g., impurities or interface defects. What remains are carrier-phonon interactions, which play an important role at room temperature \cite{Bai2022}. They therefore require a special treatment. 

\textit{Ab initio} methods, in particular density functional theory (DFT), have proven invaluable to predict the carrier transport properties of solids in the presence of interactions with phonons. These models involve combining electronic band structures \cite{Kresse1996,Giannozzi2009} with phonon dispersions \cite{Togo2015,Baroni2001} to obtain the required electron–phonon (ep) interaction elements \cite{Frederiksen2007,Gunst2016,Ponce2016,Zhou2021}. The success of these \textit{ab initio} techniques at evaluating intrinsic carrier transport in bulk materials has led to the exploration of phonon-limited mobility in a variety of 2D semiconductors \cite{Kaasbjerg2012, Li2015, Li2013, Zhao2018, Pilotto2022, Sohier2018, Gaddemane2021, Zhang2014, Huang2016, Jin2014, Guo2019, Ponce2023}. It should be emphasized that these studies on 2D materials often report a broad spectrum of mobility values, revealing a lack of consensus and frequent discrepancies with experimental data. Several factors can explain the large range of mobility values, the different DFT settings that are used being the most important one.

The knowledge of the phonon-limited mobility provides insight into the potential of a given material as transistor channel, but it only tells one side of the story. To assess the suitability of integrating a material into next-generation logic devices, the scope of inquiry must be broadened. Ultimately, what matters is the influence of carrier-phonon interactions on the current that flows through the transistor. While previous 2D device simulations have included electron-phonon interactions \cite{Szabo2015EP,Lee2019,Fiore2022}, they have often relied on simplified, less accurate models where only a limited portion of these interactions is retained. The absence of a unified model that allows for the inclusion of electron-phonon scattering both in materials and device calculations lays the groundwork for the current study. It aims to link the materials and device properties within a single framework to enable realistic investigations of transistors with a 2D channel material.  

We introduce a fully atomistic approach to compute electron-phonon interactions that facilitates the incorporation of materials properties into device calculations with focus on 2D materials \cite{Backman2022}. The method exploits DFT and maximally localized Wannier functions (MLWF) to extract material data that can be fed either to a linearized Boltzmann transport equation (LBTE) or non-equilibrium Green’s Function (NEGF) solver, delivering a holistic perspective on phonon-limited transport properties. Through the deployment of this approach, the impact of electron-phonon scattering on the functionality of 2D devices can be accurately determined, enhancing our understanding of the mechanisms limiting the performance of these compounds. Furthermore, mobility serves as a benchmark to validate and cross-correlate results obtained with the LBTE and NEGF methods. Such comparisons ensure the robustness and reliability of the derived insights. 

The paper is organized as follows. Section \ref{sec:method} presents the developed modeling techniques, starting with the approach to calculate electron-phonon coupling elements. This section also outlines the electronic structure method we adopted and elaborates on how electron-phonon scattering is treated within the LBTE and NEGF frameworks. Section \ref{sec:Results} concentrates on the results of our research. First, electronic and phononic dispersions are reported for various TMDCs before the mobility values obtained via both LBTE and NEGF methodologies are compared with each other. Monolayer MoS$_2$ is chosen as testbed. This section also introduces device simulations, highlighting the role of electron-phonon scattering on the current magnitude and on its distribution. The paper is summarized and conclusions are drawn in Section \ref{sec:Con}.

\section{Methods}\label{sec:method}

In this section the theoretical basis and computational frameworks pivotal to this study are laid out. We start by constructing an atomistic electron-phonon scattering model, specifically tailored for electronic transport calculations. A transformation from a plane-wave basis to localized electronic states is required for that purpose. Special attention is paid to the periodicity and confinement of the dimensional degrees of freedom pertinent to 2D materials and devices. Subsequent sections address the application of the LBTE and NEGF to calculate the transport properties of the considered systems. The emphasis is set on the integration of the developed electron-phonon scattering model. The goal here is to provide detailed explanation of all ingredients entering our approach, setting the stage for the upcoming discussion on our findings and their broader implications.

\subsection{Electronic and Vibrational Structure Methods}

Our method to compute electron-phonon coupling elements is outlined here. It relies on the harmonic approximation, DFT, MLWF \cite{Marzari1997}, and the frozen-phonon approach \cite{Togo2015}. Starting from the Born-Oppenheimer adiabatic approximation, which allows for the separation of electronic and vibrational time scales \cite{Horsfield2006}, the total Hamiltonian of the system of interest can be decomposed into three distinct components
\begin{dmath}
        \mathbf{\hat{H}} = \mathbf{\hat{H}}_{e}^0 + \mathbf{\hat{H}}_{p}^0 + \mathbf{\hat{H}}_{ep}.
\label{eq:Htot}
\end{dmath}
In Eq.~(\ref{eq:Htot}), $\mathbf{\hat{H}}_{e}^0$ is the single-particle mean-field Hamiltonian. It governs the behavior of electrons within a fixed ionic lattice. Conversely, $\mathbf{\hat{H}}_{p}^0$ is the Hamiltonian describing independent, non-interacting phonons. Finally, $\mathbf{\hat{H}}_{ep}$ denotes the electron-phonon Hamiltonian, which captures the interaction between these two populations.

\subsubsection{Electron-Phonon Coupling}\label{sec:ep}

The Hamiltonian encompassing the electron-phonon interactions in a periodic system can be generally written as \cite{Mahan2000,Wacker2002}:
\begin{dmath}
\mathbf{\hat{H}}_{ep} = \sum_{mn\mathbf{k}} \sum_{\lambda\mathbf{q}} M_{mn}^{\lambda}(\mathbf{k},\mathbf{q})\hat{\mathbf{c}}_{m\mathbf{k+q}}^{\dagger}\hat{\mathbf{c}}_{n\mathbf{k}} [\hat{\mathbf{b}}_{\lambda\mathbf{-q}}^{\dagger} + \hat{\mathbf{b}}_{\lambda\mathbf{q}}],
\label{eq:Hep}
\end{dmath}
where $\hat{\mathbf{c}}_{m \mathbf{k+q}}^{\dagger}(\hat{\mathbf{c}}_{n \mathbf{k}})$ and $\hat{\mathbf{b}}_{\lambda\mathbf{-q}}^{\dagger}(\hat{\mathbf{b}}_{\lambda\mathbf{q}})$ are the electron and phonon creation (annihilation) operators, respectively. The indices $m$ and $n$ correspond to the basis functions of the electrons, while $\lambda$ represents the phonon band index. The momentum vectors of electrons are indicated by $\mathbf{k}$, while those of phonons are given by $\mathbf{q}$. To understand the form of Eq.~(\ref{eq:Hep}), we should go back to the Hamiltonian of the electronic system, $\mathbf{\hat{H}}_e$. Because electrons operate on a far shorter timescale than heavy nuclei, the adiabatic approximation suggests that the nuclear coordinates have a parametric influence on the electronic Hamiltonian, i.e., $\mathbf{\hat{H}}_e = \mathbf{\hat{H}}_e(\mathbf{Q})$, where $\mathbf{Q} = \boldsymbol{\tau} - \boldsymbol{\tau}^0$ is the displacement of an atom at position $\boldsymbol{\tau}$ from its static ionic lattice configuration $\boldsymbol{\tau}^0$. To account for the perturbation $\mathbf{Q}$, the electronic Hamiltonian can be expanded in a Taylor series around the equilibrium positions of the atoms. In the harmonic approximation, only the first order terms are retained \cite{Frederiksen2007},
\begin{dmath}
\mathbf{\hat{H}}_e \approx {\mathbf{\hat{H}}_e^0 +  \sum_{I \eta} \braket{\hat{\Psi}|\frac{\partial \mathbf{\hat{H}}_{e}^0}{\partial Q_{I\eta}}|\hat{\Psi}} \hat{\mathbf{Q}}_{I\eta}  = \mathbf{\hat{H}}_e^0 + \mathbf{\hat{H}}_{ep}}.
\label{eq:purH}
\end{dmath}
In this expression, the first term corresponds to the equilibrium electronic Hamiltonian $\mathbf{\hat{H}}_{e}^0$ in Eq.~(\ref{eq:Htot}). The second term, which includes the derivative of the electronic Hamiltonian operator with respect to the displacement of atom $I$ along the Cartesian coordinate $\eta$, represents the interaction between electrons and phonons. It is projected onto a basis made of the wave functions $\ket{\hat{\Psi}}$. The displacement of the atoms from their equilibrium position due to lattice vibrations is given by the quantum operator \cite{Wacker2002,Mahan2000}
\begin{dmath}
\hat{\mathbf{Q}}_{I\eta} = \sum_{\gamma \lambda \mathbf{q}}\sqrt{\frac{\hbar}{2
N_{\mathbf{q}}m_I\omega_{\lambda\mathbf{q}}}}f_{\lambda\mathbf{q}}^{I\eta} e^{i\mathbf{q}\cdot\mathbf{R}_\gamma}[\hat{\mathbf{b}}_{\lambda\mathbf{-q}}^{\dagger}+\hat{\mathbf{b}}_{\lambda\mathbf{q}}],
\label{eq:Q}
\end{dmath}{}
where $N_{\mathbf{q}}$ represents the number of $\mathbf{q}$-points, the sum over $\gamma$ runs over the unit cells in the macroscopic system, $m_I$ is the mass of the displaced atom, and $f_{\lambda\mathbf{q}}^{I\eta}$ is the ionic displacement vector of mode $(\lambda,\mathbf{q})$ with frequency $\omega_{\lambda\mathbf{q}}$. It is assumed that the perturbed ion is located in a unit cell $\gamma$ that was displaced from the reference cell by a lattice vector $\mathbf{R}_{\gamma}$.
Next, we expand the wave function $\ket{\hat{\Psi}}$ in terms of a localized basis set
\begin{dmath}
\ket{\hat{\Psi}} =  \frac{1}{\sqrt{N}}\sum_{\alpha n\mathbf{k}} e^{i\mathbf{k}\cdot\mathbf{R}_{\alpha}}\ket{n,\mathbf{R}_{\alpha}}\hat{\mathbf{c}}_{n\mathbf{k}}.
\label{eq:Psi}
\end{dmath}
Here $\ket{n,\mathbf{R}_\alpha}$ is a state localized at position $\mathbf{R}_{\alpha}$ in unit cell ${\alpha}$ and $N$ is the number of of unit cells. Inserting Eqs. (\ref{eq:Q}) and (\ref{eq:Psi}) into Eq.~(\ref{eq:purH}) now gives us an expression for the electron-phonon interaction Hamiltonian 
\begin{dmath}
\mathbf{\hat{H}}_{ep} =
\frac{1}{N}\sum_{\alpha n\mathbf{k}}\sum_{\beta m\mathbf{k'}}\sum_{ I \eta \gamma \lambda \mathbf{q}}  e^{-i\mathbf{k'}\cdot\mathbf{R}_{\beta}}e^{i\mathbf{k}\cdot\mathbf{R}_{\alpha}} e^{i\mathbf{q}\cdot\mathbf{R}_\gamma} \cdot \newline
\bra{m,\mathbf{R}_\beta}  \frac{\partial \mathbf{\hat{H}}_{e}^0}{\partial Q_{I \eta}} \ket{n,\mathbf{R}_{\alpha}} \sqrt{\frac{\hbar}{2N_{\mathbf{q}}m_I\omega_{\lambda\mathbf{q}}}} f_{\lambda\mathbf{q}}^{I\eta} 
\hat{\mathbf{c}}_{m\mathbf{k'}}^{\dagger}\hat{\mathbf{c}}_{n\mathbf{k}} [\hat{\mathbf{b}}_{\lambda\mathbf{-q}}^{\dagger} + \hat{\mathbf{b}}_{\lambda\mathbf{q}}].
\label{eq:epH1}
\end{dmath}
Because of the periodicity of the system, the term containing the derivative of the Hamiltonian $\mathbf{\hat{H}}_{e}$ can be shifted to have $\mathbf{R}_{\alpha}$ as the reference unit cell, i.e., 
$$\bra{m,\mathbf{R}_\beta}  \frac{\partial \mathbf{\hat{H}}_{e}^0}{\partial Q_{I \eta}} \ket{n,\mathbf{R}_{\alpha}} = \bra{m,\mathbf{R}_\beta-\mathbf{R}_\alpha}  \frac{\partial \mathbf{\hat{H}}_{e}^0}{\partial Q_{I \eta}} \ket{n,\mathbf{R}_{\alpha}-\mathbf{R}_\alpha}.$$
By defining all lattice vectors relative to $\mathbf{R}_{\alpha}$, we can rewrite the phase factors in Eq.~(\ref{eq:epH1}) as, $$ e^{-i\mathbf{k'}\cdot\mathbf{R}_{\beta}}e^{i\mathbf{k}\cdot\mathbf{R}_{\alpha}} e^{i\mathbf{q}\cdot\mathbf{R}_\gamma} = e^{i(\mathbf{k}+\mathbf{q}-\mathbf{k'})\cdot\mathbf{R}_{\alpha}}e^{-i\mathbf{k'}\cdot\mathbf{R}_{\beta0}} e^{i\mathbf{q}\cdot\mathbf{R}_{\gamma0}},$$
where $\mathbf{R}_{\beta0} = \mathbf{R}_{\beta} - \mathbf{R}_{\alpha}$ and $\mathbf{R}_{\gamma0} = \mathbf{R}_{\gamma} - \mathbf{R}_{\alpha}$ are relative displacements with respect to $\mathbf{R}_{\alpha}$. The sum over $\alpha$ in Eq.~(\ref{eq:epH1}) can now be analytically carried out, giving rise to a factor $\sum_\alpha e^{i(\mathbf{k}+\mathbf{q}-\mathbf{k'})\cdot\mathbf{R}_{\alpha}} = N\delta_{\mathbf{k'},\mathbf{k+q}}$, which enforces momentum conservation. By applying these simplifications to Eq.~(\ref{eq:epH1}) and comparing the result with Eq.~(\ref{eq:Hep}), we can derive an expression for the electron-phonon coupling elements

\begin{dmath}
M_{mn}^{\lambda \eta}(\mathbf{k},\mathbf{q}) =
\sum_{\beta}\sum_{I \gamma}  e^{-i(\mathbf{k+q})\cdot\mathbf{R}_{\beta0}} e^{i\mathbf{q}\cdot\mathbf{R}_{\gamma0}} \cdot \newline
\bra{m,\mathbf{R}_{\beta0}}  \frac{\partial \mathbf{\hat{H}}_{e}^0}{\partial Q_{I \eta}} \ket{n,\mathbf{0}} \sqrt{\frac{\hbar}{2N_{\mathbf{q}}m_I\omega_{\lambda\mathbf{q}}}} f_{\lambda\mathbf{q}}^{I\eta},
\label{eq:M1}
\end{dmath}
where the dependence on the displacement direction ($\eta$) is explicitly retained in the expression for the coupling elements.
  
\subsubsection{Electronic Hamiltonian}\label{secEH}
\begin{figure}[h!]
  \includegraphics[width=1.0\linewidth]{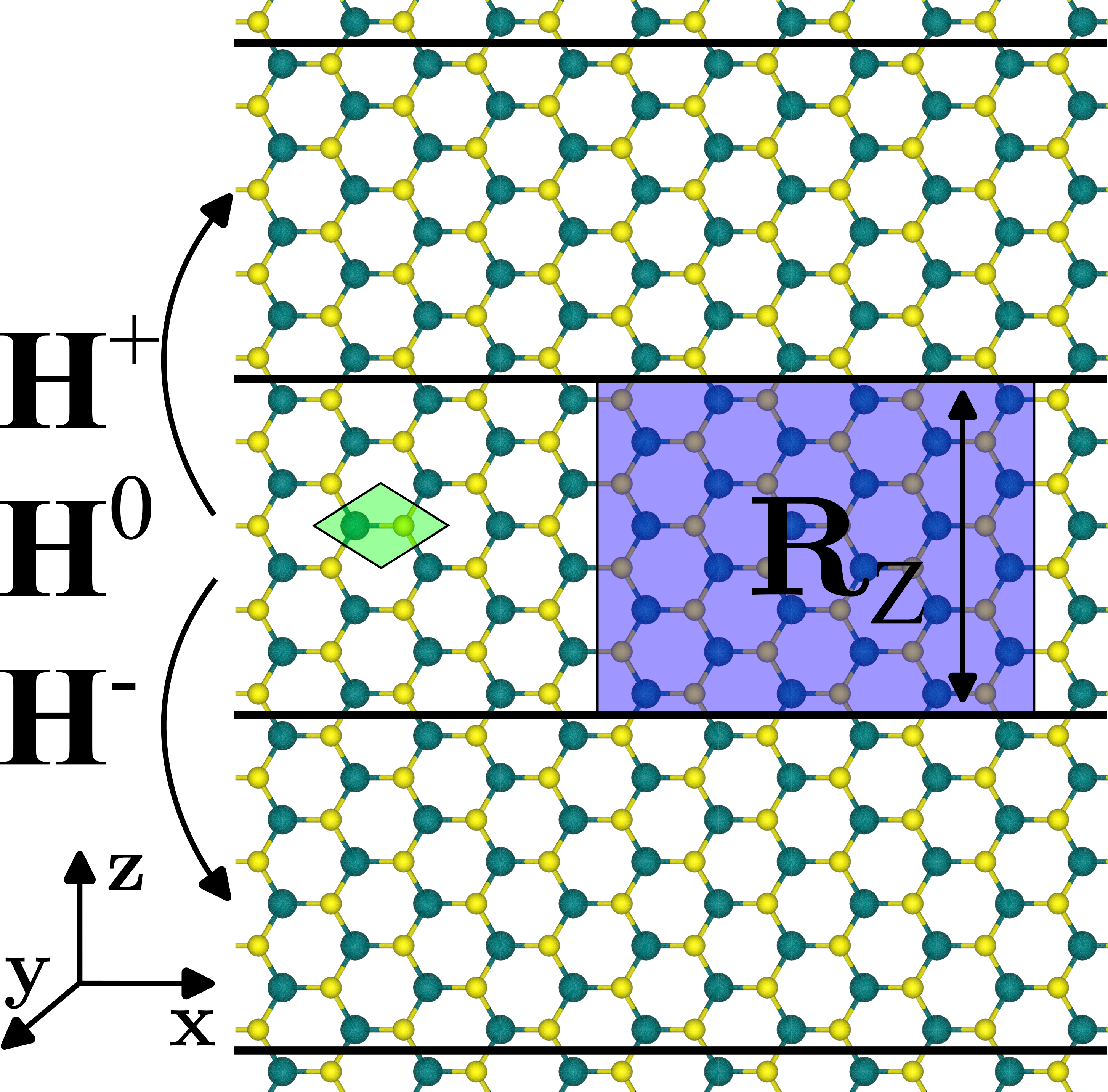}
  \caption{Top view of a MoS$_2$ monolayer constructed by replicating either the primitive cell of MoS$_2$ (green shaded area on the left) or the orthorhombic unit cell of width $R_{z}$ (blue shaded area on the right). Electrons are confined along the y-axis (out-of-plane). The primitive hexagonal unit cell of MoS$_2$ is used to calculate the electronic structure of the material, generate MLWFs, solve the LBTE system, while the orthorhombic unit cell is utilized to construct the full device Hamiltonian that enters the NEGF equations.}
  \label{fig:unitsuper}
\end{figure}

The electronic Hamiltonian $\mathbf{\hat{H}}_{e}^0$ plays a key role in electronic transport, both directly and through the electron-phonon coupling elements. Here, it is calculated from first-principles using plane-wave DFT, as implemented in the VASP package \cite{Kresse1996}. The resulting Bloch wave functions are then transformed into a set of MLWFs using the Wannier90 code \cite{Pizzi2020}
\begin{dmath}
\ket{n,\mathbf{R}} = \frac{V}{(2\pi)^3} \int_{BZ} d\mathbf{k}e^{-i\mathbf{k}\cdot \mathbf{R}} \sum_{l} \mathbf{U}_{nl}(\mathbf{k}) \ket{l,\mathbf{k}}.
\label{eq:wanfunc}
\end{dmath}{}
In Eq.~(\ref{eq:wanfunc}), the $\ket{l,\mathbf{k}}$ are the Bloch wave functions with band index $l$ and momentum $\mathbf{k}$, whereas V is the volume of the unit cell. Calculating the MLWFs primarily consists of finding the unitary transformation matrix $\mathbf{U}_{nl}(\mathbf{k})$ representing a rotation of the Bloch states that minimizes the quadratic spread of the Wannier functions. This is done through an iterative process where the electronic Bloch states projected onto atomic orbitals serve as initial guess. Once the $\mathbf{U}_{nl}(\mathbf{k})$ matrix has been determined, it can be used to transform the Bloch Hamiltonian $H_{ij}(\mathbf{k})$ into the MLWF basis, 
\begin{dmath}
H_{mn}(\mathbf{R}) = \newline \frac{V}{(2\pi)^3} \int_{BZ} d\mathbf{k}e^{-i\mathbf{k}\cdot \mathbf{R}} \sum_{i,j} 
\mathbf{U}_{mi}(\mathbf{k}) H_{ij}(\mathbf{k}) \mathbf{U}_{jn}^{\dagger}(\mathbf{k}),
\label{eq:wanH}
\end{dmath}{}
where we define $H_{mn}(\mathbf{R}) = \bra{m,\mathbf{R}} \mathbf{\hat{H}}_{e}^0\ket{n,\mathbf{0}}$. The localized nature of the Wannier functions makes the Hamiltonian matrix sparse; the interactions between Wannier functions rapidly decrease as the distance between them increases. This makes the MLWF basis suitable for transport calculations at the unit cell or device level. By associating each Wannier function with atoms located on a regular lattice, we can remap the Hamiltonian into a larger orthorhombic cell that only includes interactions with first nearest-neighbors (NN) in $\mathbf{R}$ \cite{Stieger2020}. This remapping is depicted in Fig.~\ref{fig:unitsuper}. By applying this scheme, the same DFT Hamiltonian can either be (i) interpolated onto a dense momentum grid $(k_x,k_y,k_z)$ for 3D or $(k_x,k_y)$ for 2D materials 
\begin{dmath}
    H_{mn}(\mathbf{k}) =  \sum_{\mathbf{R}}H_{mn}(\mathbf{R}) e^{i\mathbf{k}\cdot\mathbf{R}},
    \label{eq:LBTEHam}
\end{dmath}
to calculate mobilities with a LBTE solver or (ii) extended in space to correspond to the device structure considered. In the latter case, the periodicity along the transport direction is broken and the resulting system of equations is solved with NEGF. For a 2D device, with transport along the x-axis, confinement along the y-axis, and with the z-axis periodic, the device Hamiltonian becomes $k_z$-dependent $H_{mn}(k_z)$ and Eq.~(\ref{eq:LBTEHam}) reduces to
\begin{dmath}
    H_{mn}(k_z) =  H_{mn}^{0} + H^{-}_{mn}e^{-ik_zR_{z}} + H^{+}_{mn}e^{ik_zR_{z}}.
    \label{eq:deviceHam}
\end{dmath}
Here, orthorhombic unit cells of width $R_z$ along $z$ are assumed. Interactions within this unit cell are cast into the matrix $H^{0}$, while connections to the nearest-neighbor cell at $+z$ $(-z)$ form the matrix $H^{+}$ $(H^{-})$. As already mentioned, $R_z$ is chosen large enough so that only nearest-neighbor cell interactions exist \cite{Szabo2015EP}. 

\subsubsection{Hamiltonian Derivative}
Obtaining the electron-phonon coupling elements in Eq.~(\ref{eq:M1}) requires the computation of expectation values of the gradient of the Hamiltonian operator. Following the approach of \cite{Frederiksen2007}, one can bypass this direct calculation by focusing on the derivative of the Hamiltonian matrix elements and taking into account the change of basis functions \cite{Head1992}. This results in a compact expression for the sought derivative
\begin{dmath}
    \bra{m}\frac{\partial \mathbf{\hat{H}}_{e}^0}{\partial Q_{I\eta}}\ket{n} = \frac{\partial\bra{m} \mathbf{\hat{H}}_{e}^0\ket{n}}{\partial Q_{I\eta}} - 
    \\
    \left(\bra{m'} \mathbf{\hat{H}}_{e}^0\ket{n} + \bra{m} \mathbf{\hat{H}}_{e}^0\ket{n'}\right),
\label{eq:prodrule}
\end{dmath}
where $\ket{n'} \equiv \frac{\partial\ket{n}}{\partial Q_I}$ represents the shift in basis orbital $\ket{n}$ associated with the displacement $Q_{I\eta}$. The dependence on $\mathbf{R}$ in Eq.~(\ref{eq:prodrule}) has been dropped for simplicity. The first part of this expression, the gradient of the Hamiltonian matrix elements, can be evaluated in a straightforward manner by using a finite difference scheme in the localized basis,
\begin{dmath}
    \frac{\partial\bra{m} \mathbf{\hat{H}}_{e}^0\ket{n}}{\partial Q_{I\eta}} = \frac{[H_{mn}(Q_{I\eta})-H_{mn}(-Q_{I\eta})]}{2Q_{I\eta}}.
\label{eq:finite}
\end{dmath}
Here, $H_{mn}(Q_{I\eta})$ represents the Hamiltonian matrix element $\bra{m} \mathbf{\hat{H}}_{e}^0\ket{n}$ where atom $I$ has been displaced by a distance $Q$ along the direction $\eta$. To compute this derivative, a supercell is first built by repeating the primitive unit cell of the considered material along its periodic lattice vectors. Second, a collection of supercells is assembled by individually displacing the atoms situated within the central primitive unit cell. The size of the supercell must be chosen in such a way that all parameter combinations ($m,n,I,\eta$) that lead to a non-negligible derivative in Eq.~(\ref{eq:finite}) remain within the constructed domain. The number of displacements ($I,\eta$) can be reduced by limiting the displaced atoms and directions to symmetrically independent perturbations. Finally, a MLWF Hamiltonian is obtained for each displacement based on the method previously described. The localized nature of the MLWFs ensures that the derivative of the Hamiltonian is non-zero primarily in the vicinity of the displaced atom so that the sums over $\beta$ and $\gamma$ in Eq.~(\ref{eq:M1}) is limited to neighboring cells only. A key advantage of this approach resides in its ability to evaluate the derivative of the Hamiltonian operator regardless of the specificities of the selected electronic structure code, contrary to the requirements imposed by density functional perturbation theory (DFPT). Our method shows enhanced flexibility, making it compatible with various DFT codes, either through the transformation of plane-waves into Wannier functions or by directly employing a localized basis set, e.g., linear combination of atomic orbitals (LCAO) \cite{Artacho2008} or Gaussian-type orbitals (GTO) \cite{Kuehne2020}.

The second part of Eq.~(\ref{eq:prodrule}) can be expanded by introducing the projection operator
\begin{dmath}
    \sum_{kl}\ket{k}(\mathbf{S}^{-1})_{kl}\bra{l} = 1,
\label{eq:overlap}
\end{dmath}
where $\mathbf{S}$ is the orbital overlap matrix. By inserting Eq.~(\ref{eq:overlap}) into Eq.~(\ref{eq:prodrule}) we obtain a more practical form of the Hamiltonian derivative \cite{Frederiksen2007}, 
\begin{dmath}
\begin{split}
& \bra{m}\frac{\partial \mathbf{\hat{H}}_{e}^0}{\partial Q_{I}}\ket{n} = \frac{\partial\bra{m} \mathbf{\hat{H}}_{e}^0\ket{n}}{\partial Q_{I}} - \\
& \sum_{kl}\braket{m'|k}(\textbf{S}^{-1})_{kl}\bra{l}\mathbf{\hat{H}}_{e}^0\ket{n} + \bra{m}\mathbf{\hat{H}}_{e}^0\ket{k}(\textbf{S}^{-1})_{kl}\braket{l|n'}.
\end{split}
\label{eq:dHdQcorr}
\end{dmath}  
Here, $\braket{m'|k}$ and $\braket{l|n'}$ are derivatives of the orbital overlaps. They can be determined in the same way as the gradient of the Hamiltonian matrix elements in Eq.~(\ref{eq:finite}), i.e., through finite differences. The overlap matrix is calculated with the MLWFs obtained in Eq.~(\ref{eq:wanfunc}).    

\subsubsection{Phonon Frequency and Vibrational Modes}

The phonon frequencies and modes are critical ingredients of the electron-phonon coupling in Eq.~(\ref{eq:M1}). They can be obtained from the Hamiltonian $\mathbf{\hat{H}}_{p}^0$ in Eq.~(\ref{eq:Htot}), known as the dynamical matrix $\mathbf{\Phi}(\mathbf{q})$. Similar to the electronic Hamiltonian in Eq.~(\ref{eq:LBTEHam}), it is the Fourier transform of the real space dynamical matrix often referred to as the mass-scaled interatomic force constants $C_{I\eta J \xi}$
\begin{dmath}
    \Phi_{I\eta J \xi}(\mathbf{q}) = \sum_{\mathbf{R}} C_{I\eta J \xi}(\mathbf{R}) e^{i\mathbf{q}\cdot\mathbf{R}}, 
    \label{eq:fourierDyn}
\end{dmath}
where $\mathbf{R}$ points to the unit cell where atom $J$ is located. $C_{I\eta J \xi}$ represents the derivative of the Born-Oppenheimer total energy surface $E(\textbf{R})$ with respect to the atomic displacements $Q_{I\eta}$ and $Q_{J\xi}$, defined as in Eq.~(\ref{eq:finite}) \cite{Baroni2001},
\begin{dmath}
C_{I\eta J \xi} = \frac{1}{\sqrt{M_I M_J}} \frac{\partial^2E(\textbf{R})}{\partial Q_{I\eta} \partial Q_{J\xi}}.
\label{eq:FC1}
\end{dmath}{}
Since the force $F_{I\eta}$ acting on an atom $I$ along $\eta$ is readily available as an output of most electronic structure codes and because  $\partial^2E(\textbf{R}) / \left(\partial Q_{I\eta}\partial Q_{J\xi} \right) = - \partial F_{I\eta} / \partial Q_{J\xi}$, we can approximate the force constants as,   
\begin{dmath}
C_{I\eta J \xi} \approx -\frac{1}{\sqrt{M_I M_J}} \frac{[F_{I\eta}(Q_{J\xi})-F_{I\eta}(-Q_{J\xi})]}{2Q_{J\xi}}.
\label{eq:FC2}
\end{dmath}{}
This so-called frozen-phonon approach allows for the simultaneous calculation of the interatomic force constants and Hamiltonian derivatives, as described in the previous section. To avoid numerical inaccuracies related to electronic structure calculations, we enforce the acoustic sum rule and apply a numerical symmetrization of $C_{I\eta J \xi}$ \cite{Ackland1997}. The Phonopy package lends itself optimally to the construction of the displaced supercells and the subsequent calculation of the interatomic force constants \cite{Togo2023}. As for the real space MLWF Hamiltonian, the force constants can be remapped into a larger transport cell from the original DFT unit cell. Finally, the ion displacement vectors $f_{\lambda\mathbf{q}}^{I\eta}$ (phonon mode) and frequency $\omega_{\lambda \textbf{q}}$ are calculated by solving the following eigenvalue problem involving the dynamical matrix $\Phi_{I\eta J \xi}(\mathbf{q})$,
\begin{dmath}
\sum_{J \xi} \Phi_{I\eta J \xi}(\mathbf{q})f_{\lambda\mathbf{q}}^{J\xi} - \omega_{\lambda \textbf{q}}^2 f_{\lambda\mathbf{q}}^{I\eta} = 0.
\label{eq:EigDyn}
\end{dmath}

\subsection{Electronic Transport}

This section briefly summarizes the two transport methods used in this work, LBTE and NEGF. We show how the electron-phonon coupling elements that were derived in the previous section can be inserted into both formalisms to calculate phonon-limited mobility and extract electrical currents in the case of NEGF.  

\subsubsection{LBTE}

In a uniform system, under a constant, time-independent electric field, without any magnetic field, at steady-state, and with electron-phonon scattering the Boltzmann transport equation (BTE) \cite{Ziman2001,Lundstrom2002,Jacoboni2010} reduces to
\begin{dmath}
 -\frac{e\mathbf{E}}{\hbar}\frac{\partial f_{i\mathbf{k}}}{\partial \mathbf{k}} = \sum_{j \mathbf{k'} \lambda \mathbf{q}}  f_{j\mathbf{k'}}\left(1- f_{i\mathbf{k}}\right)\Gamma_{ji\mathbf{k'}}^{\lambda\mathbf{q}} - f_{i\mathbf{k}}\left(1- f_{j\mathbf{k'}}\right)\Gamma_{ij\mathbf{k}}^{\lambda\mathbf{q}}.
 \label{eq:BTE}
\end{dmath}
Here, $e$ stands for the elementary charge, $\hbar$ represents the reduced Planck's constant, and $\mathbf{E}$ denotes the electric field. The term $f_{i\mathbf{k}}$ ($f_{j\mathbf{k'}}$) refers to the electronic distribution function for a state $|i\mathbf{k}\rangle$ ($|j\mathbf{k'}\rangle$) with band index $i$ ($j$), and momenta $\mathbf{k}$ ($\mathbf{k'}$). The scattering rate, mediated by a phonon mode $\lambda$ and wave vector $\textbf{q}$, is given by $\Gamma_{ij\mathbf{k}}^{\lambda\mathbf{q}}$ ($\Gamma_{ji\mathbf{k'}}^{\lambda\mathbf{q}}$). An approximate solution to Eq.~(\ref{eq:BTE}) can be constructed by linearizing $f_{i\mathbf{k}}$ with respect to the electric field with the help of the following ansatz $f_{i\mathbf{k}} \approx f_{i\mathbf{k}}^0 + \frac{\partial f_{i\mathbf{k}}^0}{\partial E_{i\mathbf{k}}} e\mathbf{E}\cdot\mathbf{F}_{i\mathbf{k}}$, where $f_{i\mathbf{k}}^0$ is the equilibrium Fermi distribution function and $\mathbf{F}_{i\mathbf{k}}$ the carrier mean free displacement vector \cite{Li2014}. It represents the quantity to be calculated. By applying this linear ansatz to Eq.~(\ref{eq:BTE}) together with the conditions of detailed balance \cite{Kawamura1992} and momentum conservation, we arrive at the LBTE \cite{Rhyner2013} 
\begin{dmath}
    \mathbf{v}_{i\mathbf{k}} = \sum_{j \lambda \mathbf{q}} \frac{1- f_{j\mathbf{k+q}}^0}{1- f_{i\mathbf{k}}^0} \Gamma_{ij\mathbf{k}}^{\lambda\mathbf{q}} \left( \mathbf{F}_{i\mathbf{k}} - \mathbf{F}_{j\mathbf{k+q}} \right),
    \label{eq:LBTE}
\end{dmath}
where $\mathbf{v}_{i\mathbf{k}}$ is the electronic group velocity.

The electronic state $|i\mathbf{k}\rangle$ with energy $E_{i\mathbf{k}}$ is a solution of the eigenvalue problem involving the Hamiltonian depicted in Eq.~(\ref{eq:LBTEHam})
\begin{dmath}
    \sum_{n} H_{mn}(\mathbf{k})c_{i\mathbf{k}}^{n} - E_{i\mathbf{k}}c_{i\mathbf{k}}^{m} = 0,
    \label{eq:EigH}
\end{dmath}
where the $c_{i\mathbf{k}}^{n}$ are the expansion coefficients for the electron wave function. The scattering rate $\Gamma_{ij\mathbf{k}}^{\lambda\mathbf{q}}$ from an initial electronic state $|i\mathbf{k}\rangle$ into a final state $|j\mathbf{k \pm q}\rangle$ caused by phonon absorption $(+)$ or emission $(-)$ can be calculated with Fermi's Golden Rule \cite{Dirac1927}
\begin{dmath}
\Gamma_{ij\mathbf{k}}^{\lambda\mathbf{q}} = \frac{2\pi}{\hbar} \left|g_{ij}^{\lambda}(\mathbf{k},\mathbf{q})\right|^2 \left[ N^{0}_{\lambda \mathbf{q}} \delta\left( E_{i\mathbf{k}} + \hbar\omega_{\lambda \mathbf{q}} - E_{j\mathbf{k+q}} \right)  + \newline \left( 1 + N^{0}_{\lambda -\mathbf{q}} \right) \delta\left( E_{i\mathbf{k}} - \hbar\omega_{\lambda -\mathbf{q}} - E_{j\mathbf{k-q}} \right) \right],
\label{eq:scattrate}
\end{dmath}
where $N^{0}_{\lambda \mathbf{q}}$ is the equilibrium Bose-Einstein distribution, for which it holds that $N^{0}_{\lambda \mathbf{-q}} = N^{0}_{\lambda \mathbf{q}}$ and $\omega_{\lambda \mathbf{-q}} = \omega_{\lambda \mathbf{q}}$. The probability amplitude of the scattering process $g_{ij}^{\lambda}(\mathbf{k},\mathbf{q})$ is obtained from the electron-phonon coupling elements in Eq.~(\ref{eq:M1}) as
\begin{dmath}
    g_{ij}^{\lambda}(\mathbf{k},\mathbf{q}) = \sum_{mn\eta} c_{i\mathbf{k+q}}^{m\dagger} M_{mn}^{\lambda\eta}(\mathbf{k},\mathbf{q}) c_{j\mathbf{k}}^{n}
    \label{eq:M2}
\end{dmath}

Eq.~(\ref{eq:LBTE}) forms a linear system of equations whose size rapidly explodes as the number of $\mathbf{k}$/$\mathbf{q}$-points increases. Given the necessity for dense Brillouin Zone sampling to ensure precise results, solving the LBTE becomes a computationally demanding task. To address this issue, an iterative approach can be employed. We first consider a simpler solution where we neglect the $\mathbf{F}_{j\mathbf{k+q}}$ term in Eq.~(\ref{eq:LBTE}). This is referred to as the energy relaxation time approximation (ERTA) \cite{Ziman2001,Lundstrom2002}, which leads to a simple solution for Eq.~(\ref{eq:LBTE}), namely $\mathbf{F}_{i\mathbf{k}}^{\text{ERTA}} = \mathbf{v}_{i\mathbf{k}}  \tau_{i\mathbf{k}}$, with the generalized relaxation time $\tau_{i\mathbf{k}}$ defined as \cite{Li2015,Ponce2018}
\begin{dmath}
    \frac{1}{\tau_{i\mathbf{k}}} = \frac{2\pi}{\hbar} \sum_{j \lambda \mathbf{q}} \left|g_{ij}^{\lambda}(\mathbf{k},\mathbf{q})\right|^2 \cdot 
    \newline
    \left[ \left(N^{0}_{\lambda \mathbf{q}} + f_{j\mathbf{k+q}}^0 \right) \delta\left( E_{i\mathbf{k}} + \hbar\omega_{\lambda \mathbf{q}} - E_{j\mathbf{k+q}} \right) + 
    \newline  
    \left( 1 + N^{0}_{\lambda -\mathbf{q}} - f_{j\mathbf{k-q}}^0 \right) \delta\left( E_{i\mathbf{k}} - \hbar\omega_{\lambda -\mathbf{q}} - E_{j\mathbf{k-q}} \right) \right].
    \label{eq:relaxTime}
\end{dmath}
The exact solution of Eq.~(\ref{eq:LBTE}) can then be iteratively computed with $\mathbf{F}_{i\mathbf{k}}^{\text{ERTA}}$ as initial guess, i.e., $\mathbf{F}_{i\mathbf{k}}^{0} 	\equiv \mathbf{F}_{i\mathbf{k}}^{\text{ERTA}}$ \cite{Li2014}. The iterative procedure obeys the following equation 
\begin{dmath}
    \mathbf{F}_{i\mathbf{k}}^{\xi+1} = \mathbf{F}_{i\mathbf{k}}^{0} +  \tau_{i\mathbf{k}} \sum_{j \lambda \mathbf{q}}\frac{1- f_{j\mathbf{k+q}}^0}{1- f_{i\mathbf{k}}^0} \Gamma_{ij\mathbf{k}}^{\lambda\mathbf{q}}\mathbf{F}_{j\mathbf{k+q}}^{\xi} 
    \label{eq:iter}
\end{dmath}.
Convergence is achieved when the low field conductivity
\begin{dmath}
    \sigma = \frac{2e^2}{Vk_B T} \sum_{i \mathbf{k}} f_{i\mathbf{k}}^0\left(1- f_{i\mathbf{k}}^0\right) F^{\eta \xi}_{i\mathbf{k}} \mathbf{v}^{\eta}_{i\mathbf{k}}
    \label{eq:condLBTE}
\end{dmath}
does not change by more than a pre-defined criterion between two consecutive iterations $(\xi+1$ and $\xi)$. In Eq.~(\ref{eq:condLBTE}) $\eta$ is the direction of the electric field, $k_B$ Boltzmann's constant, $T$ the temperature, and the factor 2 accounts for spin degeneracy. Finally, once the iterative solution has converged the phonon-limited mobility $\mu$ is given by
\begin{dmath}
    \mu = \frac{\sigma}{e n},
    \label{eq:mob}
\end{dmath}
where $n$ is the electron density. It it calculated according to the equilibrium Fermi distribution function
\begin{dmath}
    n = \frac{2}{V} \sum_{n \mathbf{k}} f_{n\mathbf{k}}^0.
    \label{eq:cdLBTE}
\end{dmath}

The electronic group velocity, which enters Eq.~(\ref{eq:condLBTE}), is defined as \cite{Rhyner2013} 
\begin{dmath}
\mathbf{v}_{i\mathbf{k}} = \frac{1}{\hbar} \sum_{mn} c_{i\mathbf{k}}^{m\dagger} \left[\frac{\partial H_{mn}(\mathbf{k})}{\partial \textbf{k}}\right] c_{i\mathbf{k}}^{n},
\label{eq:velo}
\end{dmath}
where the $\mathbf{k}$-derivative of Hamiltonian in Eq.~(\ref{eq:LBTEHam}) can be carried out analytically \cite{Marzari2012}
\begin{dmath}
    \frac{\partial H_{mn}(\textbf{k})}{\partial k_{\eta}} = \sum_{\textbf{R}} iR_{\eta} e^{i\textbf{k}\cdot \textbf{R}}H_{mn}(\textbf{R}).
    \label{eq:dHdk}
\end{dmath}
In Eq.~(\ref{eq:dHdk}) $\eta$ indicates the direction along which the derivative is calculated.

Note that the delta function $\delta\left( E_{i\mathbf{k}} \pm \hbar\omega_{\lambda \pm\mathbf{q}} - E_{j\mathbf{k\pm q}} \right)$ in Eq.~(\ref{eq:scattrate}) is approximated as the limit of a Gaussian function whose smearing parameter approaches zero. We use an adaptive smearing method \cite{Li2015,Li2012} to determine the best-suited value of this parameter for a given $\mathbf{k}$/$\mathbf{q}$ grid.

\subsubsection{NEGF}

\begin{figure}[h!]
  \includegraphics[width=1.0\linewidth]{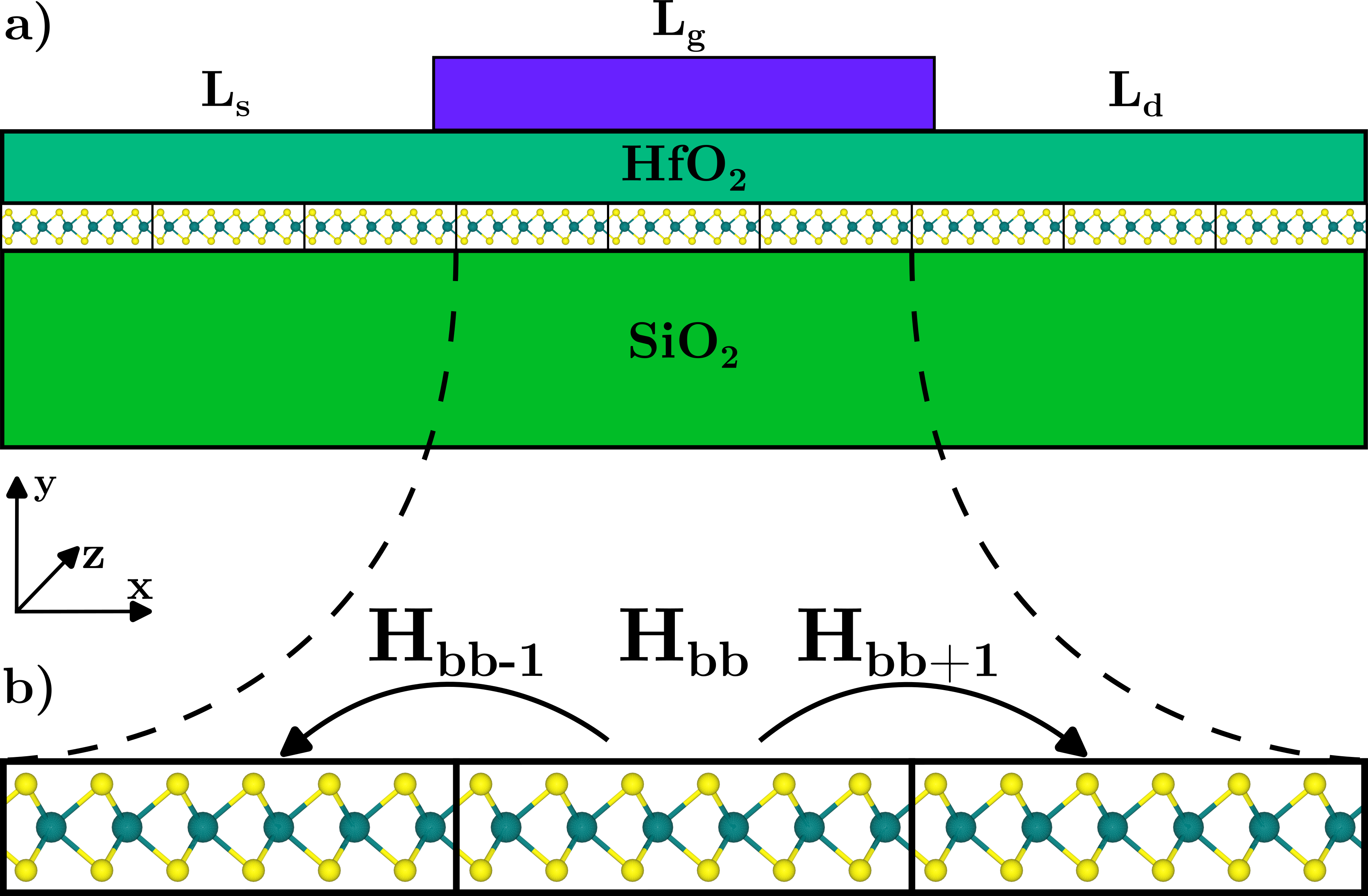}
  \caption{ (a) Schematic of a 2D single-gate transistor structure featuring a MoS$_2$ monolayer as the channel region. The lengths of the source, drain, and gate are denoted as L$_s$, L$_d$, and L$_g$, respectively. (b) Detailed view of the channel region, which is divided into orthorhombic unit cells. The corresponding Hamiltonian blocks are indicated; block b is connected with itself ($\mathbf{H}_{bb}$) as well as its next ($\mathbf{H}_{bb+1}$) and previous ($\mathbf{H}_{bb-1}$) neighbor blocks.}
  \label{eq:deviceBlocks}
\end{figure}

Next, we move beyond material properties under equilibrium conditions and target device simulation, where a system is driven out of equilibrium by the application of an external voltage. For this, the NEGF formalism has established itself as a robust computational methodology that is capable of addressing both coherent and dissipative quantum transport \cite{Danielewicz1984,Datta1990,Lake1997,Datta2000,Svizhenko2002}. Specifically, this work uses NEGF to model transistors with a 2D channel material including electron-phonon scattering through the self-consistent Born approximation \cite{Luisier2006,Luisier2009,Rhyner2013,Szabo2015EP}. In 2D systems, as in the monolayer MoS$_2$ FET depicted in Fig.~\ref{eq:deviceBlocks}, electron transport occurs along one principal axis, designated here as the x-axis. Perpendicular to this, the y-axis serves as the direction of confinement, while the z-axis is assumed periodic. In such a configuration, the following system of equations must be solved to obtain all desired observables     

\begin{equation}
  \begin{split}
    \sum_{l} \left\{\left[\mathbf{E} - \mathbf{V}(\mathbf{r_{m}})\right]\delta_{lm} - \mathbf{H}_{ml}(k_z) - \mathbf{\Sigma}_{ml}^{RB}(E,k_z) - \right. \\
    \left. \mathbf{\Sigma}_{ml}^{RS}(E,k_z)\right\}\mathbf{G}_{ln}^{R}(E,k_z) = \delta_{mn},
  \end{split}
  \label{eq:SENEGF}
\end{equation}

\begin{dmath}
    \mathbf{G}_{mn}^{\gtrless}(E,k_z) = \sum_{l_1 l_2}\mathbf{G}_{m l_1}^{R}(E,k_z)\left[\mathbf{\Sigma}_{l_1,l_2}^{\gtrless B}(E,k_z) + \\
    \mathbf{\Sigma}_{l_1,l_2}^{\gtrless S}(E,k_z)\right]\mathbf{G}_{l_2 n}^{A}(E,k_z).
    \label{eq:Ggrtless}
\end{dmath}

The indices $l$, $m$, and $n$ refer to the atomic positions $\mathbf{r}_l$, $\mathbf{r}_m$, and $\mathbf{r}_n$, respectively. The diagonal matrices $\mathbf{E}$ and $\mathbf{V}(\mathbf{r}_m)$ contain the injection energy $E$ and the self-consistent electrostatic potential V at position $\mathbf{r}_m$. The size of both matrices is $p_m \times p_m$, where $p_m$ is the number of Wannier functions located on the $m^{\text{th}}$ atom. The block Hamiltonian $ \mathbf{H}_{mn}(k_z)$ is expressed in the MLWF basis constructed according to Eq.~(\ref{eq:deviceHam}). Its entries are the matrix elements between atoms with index $m$ and $n$ and its size is $p_m \times p_n$. The scattering $\mathbf{\Sigma}_{mn}^{RS}(E,k_z)$ and boundary $\mathbf{\Sigma}_{mn}^{RB}(E,k_z)$ retarded self-energies together with the retarded, $\mathbf{G}_{mn}^{R}(E,k_z)$, advanced, $\mathbf{G}_{mn}^{A}(E,k_z) = \mathbf{G}_{mn}^{R\dagger}(E,k_z)$, lesser, $\mathbf{G}_{mn}^{<}(E,k_z)$, and greater, $\mathbf{G}_{mn}^{>}(E,k_z)$, Green's functions are also of size $p_m \times p_n$. These quantities must be solved for each injection energy $E$ and transverse momentum $k_z$.

 The boundary self-energies \cite{Luisier2006} which are different from 0 only when atoms $m$ and $n$ are directly connected to the device contacts can be computed through decimation techniques \cite{Sancho1985} or eigenvalue problems \cite{Calderara2015}. Their scattering counterparts, $\mathbf{\Sigma}_{mn}^{S}(E,k_z)$, are full matrices and consist here of electron-phonon interactions. They are obtained via perturbation theory by expanding the Green's function to the second order in the electron-phonon interaction Hamiltonian in Eq.~(\ref{eq:epH1}), applying Wick's theorem, and writing down the corresponding Dyson equation \cite{Mahan2000,Danielewicz1984,Mattuck1992}. The resulting general form of the electron-phonon scattering self-energy in the time domain is
\begin{dmath}
    \mathbf{\Sigma}_{mn}^{ep}(tt',k_z) = i\hbar \sum_{l_1 l_2 \eta_1 \eta_2}\sum_{\lambda q_z} \mathbf{M}_{m l_1}^{\lambda \eta_1}(k_z-q_z,q_z) \cdot
    \\
    \mathbf{G}_{l_1 l_2}(tt',k_z-q_z)\cdot \mathbf{M}_{l_2 n}^{\lambda \eta_2}(k_z,-q_z)  \mathbf{D}^{\lambda}(tt',q_z).
    \label{eq:SigEPtime}
\end{dmath}
In this expression, the elements of the electron-phonon coupling blocks $\mathbf{M}_{mn}^{\lambda\eta}(\mathbf{k},\mathbf{q})$ are defined as in Eq.~(\ref{eq:M1}) and $\mathbf{D}^{\lambda}(\mathbf{q},tt')$ represents the phonon Green's function. Applying Langreth theorem, replacing the phonon Green's function by its unperturbed form at equilibrium, and moving to steady-state conditions leads to the following energy-dependent greater and lesser self-energy \cite{Langreth1976,Mahan2000}
\begin{dmath}
    \mathbf{\Sigma}_{mn}^{\gtrless ep}(E,k_z) = \sum_{l_1 l_2 \eta_1 \eta_2}\sum_{\lambda q_z} \mathbf{M}_{m l_1}^{\lambda \eta_1}(k_z-q_z,q_z) \cdot
    \\
    \{N^{0}_{\lambda q_z} \mathbf{G}_{l_1l_2}^{\gtrless}(E\pm \hbar \omega_{\lambda q_z},k_z-q_z) +
    \\
    [N^{0}_{\lambda q_z} + 1]\mathbf{G}_{l_1l_2}^{\gtrless}(E\mp \hbar \omega_{\lambda q_z},k_z-q_z)\} \cdot
    \\
    \mathbf{M}_{l_2 n}^{\lambda \eta_2}(k_z,-q_z),
    \label{eq:SigEPEnergy}
\end{dmath}
where $N^{0}_{\lambda q_z}$ is the equilibrium Bose-Einstein distribution of phonons with frequency $\omega_{\lambda q_z}$. By employing the greater and lesser electron-phonon scattering self-energies $\mathbf{\Sigma}_{mn}^{\gtrless ep}(E,k_z)$, we can derive an expression for the retarded scattering self-energy $\mathbf{\Sigma}_{mn}^{RS}(E,k_z)$ in Eq.~(\ref{eq:SENEGF})

\begin{dmath}
    \mathbf{\Sigma}_{mn}^{RS}(E,k_z) = \frac{1}{2}\left( \mathbf{\Sigma}_{mn}^{>ep}(E,k_z) -  \mathbf{\Sigma}_{mn}^{<ep}(E,k_z) \right) + 
    \\
    i\mathcal{P} \int \frac{dE'}{2\pi} \frac{( \mathbf{\Sigma}_{mn}^{>ep}(E,k_z) -  \mathbf{\Sigma}_{mn}^{<ep}(E',k_z))}{E-E'},
    \label{eq:SigRet}
\end{dmath}
where $\mathcal{P}$ denotes the Cauchy principal integral \cite{Frey2008}.

Because of the interdependence between $\mathbf{G}_{mn}^{\gtrless}(E,k_z)$ and $ \mathbf{\Sigma}_{mn}^{\gtrless ep}(E,k_z)$, Eqs.  (\ref{eq:SENEGF})-(\ref{eq:Ggrtless}) and (\ref{eq:SigEPEnergy})-(\ref{eq:SigRet}) must be solved self-consistently within the Born approximation. The convergence of the loop between the Green's functions and scattering self-energies is verified by monitoring the variations of the current $I_d$ and carrier density $n(\mathbf{r}_l)$ between consecutive iterations \cite{Luisier2009,Stieger2020}. The said current $I_d$, calculated as
\begin{equation}
    \begin{aligned}
        I&_{d,b \to  b+1} = \\
        & \frac{2e}{\hbar}\sum_{k_z}\sum_{m\in b}\sum_{n \in b+1} \int \frac{dE}{2\pi} \text{tr}\left\{ \mathbf{H}_{mn}(k_z)\mathbf{G}_{nm}^{<}(E,k_z) - \right. \\
        & \left. \mathbf{G}_{mn}^{<}(E,k_z)\mathbf{H}_{nm}(k_z)\right\},
    \end{aligned}
    \label{eq:IdNEGF}
\end{equation}
represents the flow of electrons from a block of atoms in the orthorhombic cell $b$ to its neighboring cell $b+1$. The position indices, $m$ and $n$, therefore encompass all atoms within the connected cells, as exemplified in Fig.~\ref{eq:deviceBlocks}. The trace operator "tr" runs over the orbitals of atoms $m$ and $n$. Finally, the factor 2 accounts for spin. The carrier density
\begin{dmath}
    n(\mathbf{r}_l) = -2i\sum_{k_z}\int \frac{dE}{2\pi}tr\left\{ \mathbf{G}_{ll}^< (E,k_z) \right\},
    \label{eq:cdNEGF}
\end{dmath}
at atomic position $\mathbf{r}_l$ is the trace of the lesser Green's Function block corresponding to this atom.

As in the LBTE case, the phonon limited mobility can be derived from the electrical conductivity $\sigma$ through Eq.~(\ref{eq:mob}). This can be done with the dR/dL method \cite{Rim2002}, where $\sigma$ is obtained by extracting the channel resistivity $\rho$ 
\begin{dmath}
    \rho = {\frac{dR}{dL}  = \frac{\Delta V}{\Delta L}\left( \frac{1}{I_d(L+\Delta L)} - \frac{1}{I_d(L)}. \right)}.
    \label{eq:resNEGF}
\end{dmath}
The dR/dL method requires calculating the drain current $I_d(L)$ for samples of different lengths $L$. The simulations are carried out on structures that have a uniform charge density $n$, a small bias difference $\Delta V = 1$ mV between the contacts, and that include electron-phonon scattering.       

 \begin{figure}[h!]
  \includegraphics[width=1.0\linewidth]{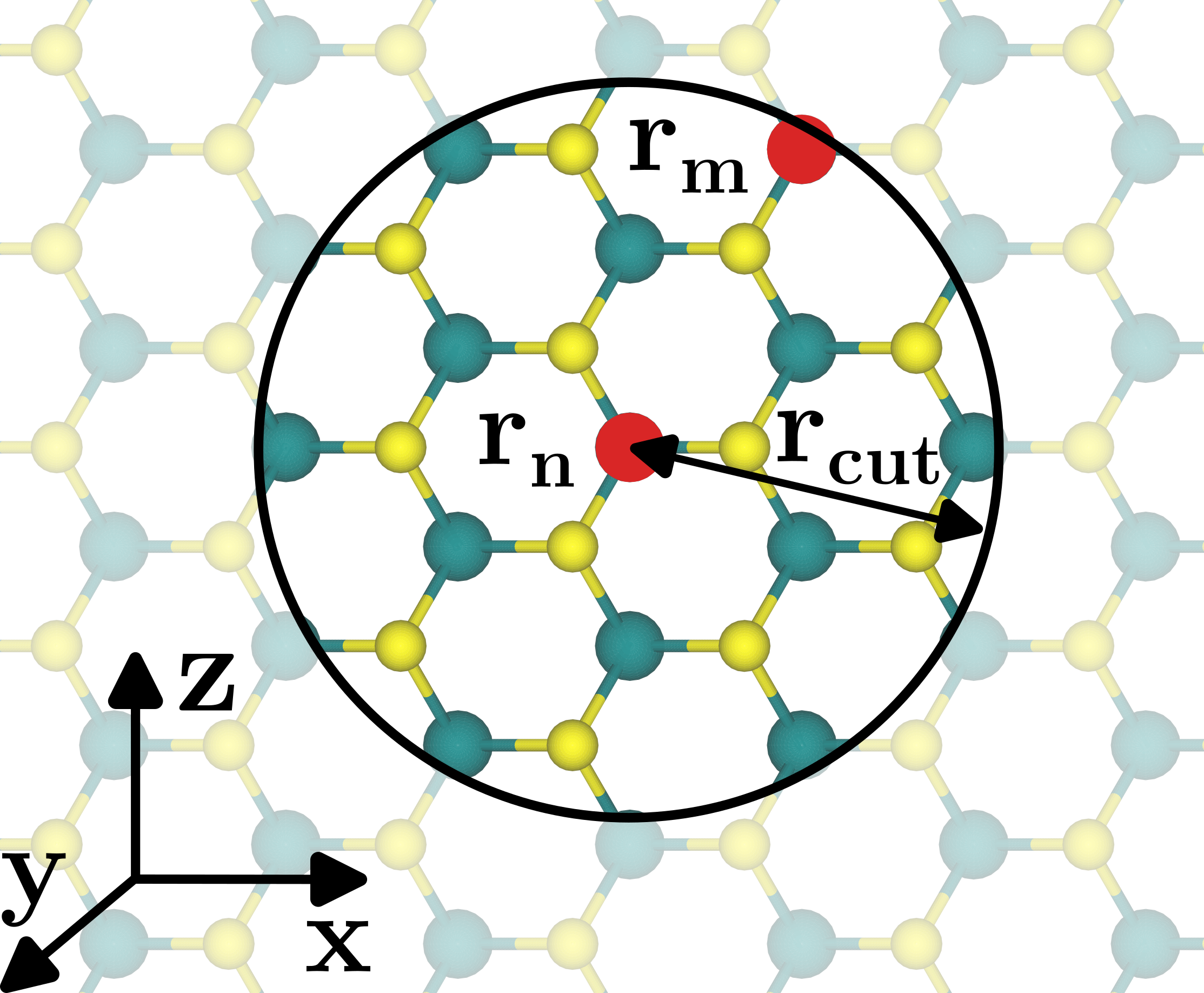}
  \caption{ Top view of a MoS$_2$ monolayer illustrating the interaction range considered when constructing the scattering self-energies $\mathbf{\Sigma}_{mn}^{ep}(E,k_z)$. Interactions between atoms $n$ and $m$ (highlighted in red) are included if their distance is smaller than a pre-defined cut-off radius $r_{cut}$—that is, if  $|r_n - r_m| < r_{cut}$. For distances exceeding this cut-off radius the self-energy entries $\mathbf{\Sigma}_{mn}^{ep}(E,k_z)$  are set to zero.} 
  \label{fig:rcut}
\end{figure}

To minimize the computational burden associated with the investigation of devices containing a large number of atoms, approximations must be applied to the calculation of the Green's functions and scattering self-energies. First, to obtain the retarded scattering self-energy in Eq.~(\ref{eq:SigRet}), the Cauchy principal integral term coupling all energies together is neglected. Since it only contributes to energy renormalization and not to relaxation or phase breaking events, leaving it out is not expected to significantly alter the device behavior, as demonstrated in previous studies \cite{Svizhenko2003,Frey2008}. Second, the phonon momentum points $\mathbf{q}$ are restricted to the $\Gamma$-point ($\mathbf{q}=\mathbf{0}$) of the orthorhombic transport cell. This simplification allows us to perform the sum over the displaced atoms ($I$ and $\gamma$), directly in Eq.~(\ref{eq:M1}). Consequently, the electron-phonon coupling element "only" depends on 5 ($\lambda,\textbf{k},m,n,\eta$) instead of 7 variables, thus significantly reducing the memory requirements and the time to evaluate Eq.~(\ref{eq:SigEPEnergy}). To compensate for the omission of the $\textbf{q}$ dependence, a large orthorhombic transport cell is chosen (R$_{z} = 2.5$ nm in Fig.~\ref{fig:unitsuper}) so that most band structure features are folded back to the $\Gamma$-point. Note that, while the number of phonon momentum points is limited, all phonon modes are taken into account individually, preserving the coupling to all phonon energies $\hbar\omega_{\lambda \mathbf{q}}$ at the $\Gamma$-point. Finally, it is not possible to treat the scattering self-energies as full matrices in large structures with thousands of atoms. A cut-off radius $r_{cut}$ is therefore introduced in Eq.~(\ref{eq:SigEPEnergy}) so that $|r_n - r_m| < r_{cut}$, as illustrated in Fig.~\ref{fig:rcut}. A convergence analysis with respect to $r_{cut}$ and comparisons with LBTE (see next section and Appendix A) indicate that such an approach is sufficient to obtain meaningful results. No scaling of the electron-phonon interactions is needed, contrary to what was proposed in \cite{Lee2019}. Despite all these simplifications and although the calculation of Eq.~(\ref{eq:SigEPEnergy}) was ported to GPUs, the computational intensity of such simulations remains gigantic, thus limiting the number of samples that could be considered.

\section{Simulation Results}\label{sec:Results}
In this section, our simulation results are presented, starting with the lattice and band structure of selected 2D materials and Silicon. We then analyze their mobility, using both LBTE and NEGF (only for MoS$_2$). Finally, we calculate the electrical current flowing through a MoS$_2$ transistor and assess the importance of employing an accurate electron-phonon scattering model.

Each electronic structure calculation was performed within the generalized gradient approximation (GGA) of Perdew, Burke, and Ernzerhof (PBE) \cite{Perdew1996}, using projector augmented wave (PAW) pseudopotentials \cite{blochl1994,Kresse1999}. A plane-wave cutoff energy of 500 eV is applied, while the total energy convergence criterion is set to less than $10^{-10}$ eV. The unit cells are relaxed to ensure that the residual forces acting on each atom are bellow $10^{-8}$ eV/$\text{Å}$. During this process, both the lattice constant and atomic positions are allowed to change. Gaussian smearing with a smearing width of 0.05 eV is employed. For the TMDCs, the electronic structure calculation is performed on a $27 \times 27 \times 1$ Monkhorst-Pack $\textbf{k}$-point grid, while a $15 \times 15 \times 15$ grid is used for Silicon. Spin-orbit coupling (SOC) is neglected. In the case of the TMDCs, a vacuum space of 20 $\text{Å}$ is used along the y-axis (direction of confinement in Fig \ref{fig:unitsuper}) to minimize interactions with periodic images. Displacement calculations are performed with $9 \times 9 \times 1$ TMDC supercells and $5 \times 5 \times 5$ Silicon supercells. A displacement distance of 0.01 $\text{Å}$ is employed to calculate all derivatives (Hamiltonian and forces). The Wannier Hamiltonians are obtained from the DFT results by considering 5 d-like orbitals on the transition metal atom and three p-like orbitals on each of the chalcogen atoms, while sp3-like orbitals are retained for each Silicon atom as initial guesses for the minimization procedure. During the Wannier optimization process a frozen-energy window \cite{Pizzi2020}, located 1 eV above the conduction band edge and 1 eV bellow the valence band edge, is defined to ensure that all states relevant for transport calculations are accurately represented by the tight-binding-like Wannier Hamiltonian.      

\subsection{Lattice and Electronic Band Structures}
It has been observed that small changes in simulation parameters, in particular DFT settings, can lead to significant differences in calculated mobilities \cite{Gaddemane2021}. It is therefore important to precisely report the crystal lattice and electronic band structures that enter the developed modeling framework to ensure meaningful comparisons with other studies.  

Starting with the crystal structure, the lattice constants and the layer thickness of the considered TMDCs, obtained through lattice relaxation, are summarized in Table \ref{tab:structureTable}. The layer thickness is measured as the distance between the top and bottom chalcogen atoms. For MoS$_2$, our calculated lattice constant is equal to 3.183 \text{\AA}, which agrees well with values reported by other groups utilizing VASP \cite{Klinkert2020,Rawat2018} or Quantum ESPRESSO \cite{Mounet2018}. A similar agreement is observed for the three other TMDCs analyzed here. It should however be noted that several studies based either on plane-wave or local-orbital methods report somewhat smaller lattice constants, ranging between 3.13 and 3.14 \text{\AA} for MoS$_2$ \cite{Kaasbjerg2012,Li2013,Zhao2018,Zhang2014,Huang2016,Jin2014}. The experimentally determined lattice constant of bulk MoS$_2$ is positioned in between and is equal to 3.15 \text{\AA} \cite{Wakabayashi1975}. For the remaining TMDCs, we calculated lattice constants of 3.318 \text{\AA} for MoSe$_2$, 3.182 \text{\AA} for WS$_2$, and 3.316 \text{\AA} for WSe$_2$. Mirroring the trend identified earlier, several studies report comparatively shorter lattice constants, e.g., 3.26-3.27 \text{\AA} for MoSe$_2$ \cite{Zhang2014,Huang2016,Jin2014} and 3.25-3.26 \text{\AA} for WSe$_2$ \cite{Zhang2014,Huang2016,Jin2014}. The range of reported values is broader for WS$_2$, with lattice constants going from 3.10 to 3.20 \text{\AA} \cite{Zhang2014,Huang2016,Jin2014}. Matching lattice parameters do not necessarily translate into identical layer thicknesses, which directly influence the vibrational modes of the material. For instance, a comparison of our results with those of \cite{Mounet2018} reveals a discrepancy of only 0.13\% in the lattice constant of MoS$_2$, whereas the layer thickness diverges by 1.6\%, a one order of magnitude larger difference.

In the case of Silicon, our calculations produce a lattice constant of 5.43 \text{\AA}. This result is in agreement with the experimentally measured value of this semiconductor \cite{SzeNg2007} and in the range of other simulation works, e.g., \cite{Ponce2018} found values between 5.40 and 5.47 \text{\AA}, depending on the exchange and correlation functional used in the DFT calculations. 

\newcolumntype{C}{>{\centering\arraybackslash}X}
\begin{table}[h!]
\begin{center}
\begin{tabularx}{\linewidth}{CCCCC}
\toprule
Material & Lattice Const. & Layer Thick. & $E_{BG}$ & $\Delta E_{KQ}$ \\
 & ($\text{Å}$) & ($\text{Å}$) & (eV) & (meV)\\
\hline
MoS$_2$     & 3.183    & 3.127 & 1.67  & 269 \\
MoSe$_2$    & 3.318    & 3.337 & 1.44  & 169 \\
WS$_2$      & 3.182    & 3.139 & 1.81  & 211 \\
WSe$_2$     & 3.316    & 3.354 & 1.54  & 131 \\
Si          & 5.430    & -     & 0.57  & -   \\
\bottomrule 
\end{tabularx}
\end{center}
\caption{Lattice and band structure parameters of selected TMDCs and Silicon. The lattice constant and distance between the upper and lower chalcogen atoms in the 2D TMDC layers are reported in the second and third columns, respectively. The variable $E_{BG}$ refers to the electronic band gap and $\Delta E_{KQ}$ to the energy separation between the $K$ and $Q$ valleys located in the conduction band of the TMDCs.}
\label{tab:structureTable}
\end{table}

\begin{figure*}[t]
  \includegraphics[width=1.0\linewidth]{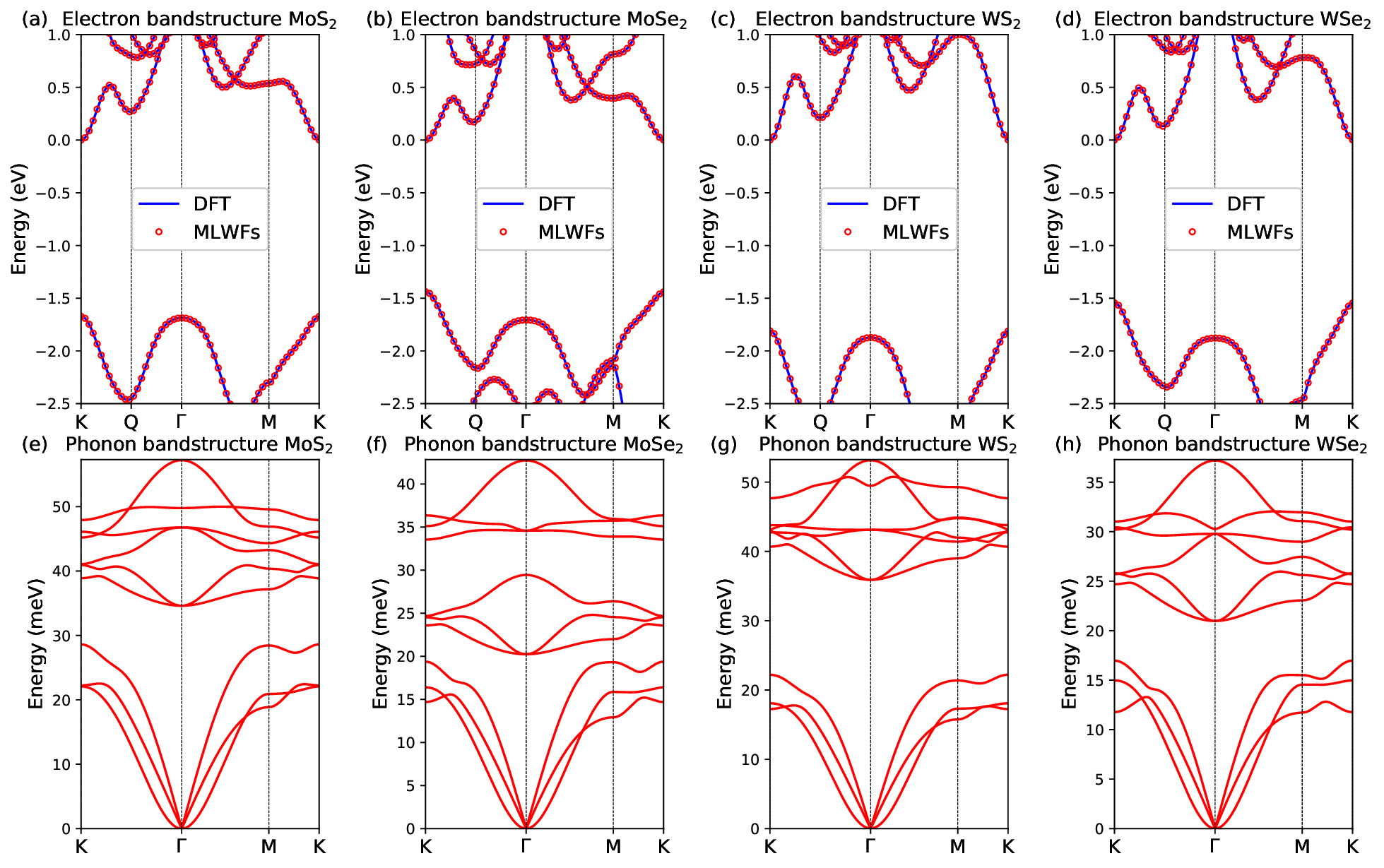}
  \caption{Electronic (a-d) and phonon (e-h) band structure of monolayer MoS$_2$, MoSe$_2$, WS$_2$, and WSe$_2$, respectively. In (a-d), the plane-wave DFT (solid lines) and MLWF (circles) band structures are compared to each other. The position of the second conduction band minimum at $Q$ is indicated along the high symmetry path. The zero energy level is set to the conduction band minimum of each material.}
  \label{fig:TMDBS}
\end{figure*}

Moving on to the electron and phonon dispersion, Fig.~\ref{fig:TMDBS} plots the electronic and phonon band structures of the selected monolayer TMDCs, while Table \ref{tab:structureTable} lists the extracted band gaps $(E_{BG})$ and the energy separation between the $K$ and $Q$ valleys $(\Delta E_{KQ})$. The exact location of the $Q$ valley is marked along the high symmetry path of the electronic band structure. The accuracy of the Wannier transformations is demonstrated by comparing the MLWF and DFT band structures. All monolayer TMDCs are found to be direct-gap semiconductors, as expected, with band gaps ranging between 1.44 and 1.81 eV. Focusing on MoS$_2$, we obtain a band gap of 1.67 eV, in agreement with \cite{Klinkert2020,Rawat2018,Mounet2018}, but slightly below the experimental band gap of 1.8 eV \cite{Mak2010}. Several simulation studies reported band gaps closer to that of experimental measurements \cite{Kaasbjerg2012,Gunst2016,Li2013,Zhao2018,Zhang2014,Huang2016,Jin2014}, but also a smaller lattice constant. A previous study \cite{Szabo2015EP} using the same simulation code, pseudopotentials, and exchange and correlation functional as here obtained a close to experimental band gap by fixing the lattice constant to that of bulk MoS$_2$. This trick also resulted in a reduction of $\Delta E_{KQ}$, a critical parameter for transport, to 48 meV, far below the 269 meV of the present study. It should however be noted that the band gap of MoS$_2$ remains an open issue. The accepted value of 1.8 eV corresponds to the optical band gap. The latter can be reproduced by DFT simulations when introducing GW corrections and accounting for excitonic effects \cite{Qiu2013}.

The $K$-$Q$ valley separation, $\Delta E_{KQ}$, in contrast to the band gap, directly influences the mobility of TMDCs through intervalley scattering. While we will delve deeper into this effect in Section \ref{sec:mob}, preliminary observations are made here. As already mentioned, $\Delta E_{KQ}$ is highly sensitive to even small variations in simulation parameters, giving rise to a large range of reported values in literature: 70-310 meV for MoS$_2$ \cite{Kaasbjerg2012,Li2015,Li2013,Zhao2018,Pilotto2022,Sohier2018,Mounet2018,Gaddemane2021,Zhang2014,Huang2016,Jin2014}, 28-155 meV for MoSe$_2$ \cite{Mounet2018,Gaddemane2021,Zhang2014,Huang2016,Jin2014}, 22-210 meV for WS$_2$ \cite{Sohier2018,Mounet2018,Gaddemane2021,Zhang2014,Huang2016,Jin2014}, and 16-124 meV for WSe$_2$ \cite{Sohier2018,Mounet2018,Gaddemane2021,Zhang2014,Huang2016,Jin2014}. It can generally be inferred that methods employing the local density approximation (LDA) tend to predict a smaller valley separation than GGA-PBE. Sensitivity to spin-orbit coupling has been shown to be TMDC-dependent. For instance, in WSe$_2$, the value of $\Delta E_{KQ}$ drops from 117 meV to 41 meV when spin-orbit coupling is taken into account. In MoSe$_2$, the inclusion of spin-orbit coupling does not significantly affect $\Delta E_{KQ}$ \cite{Gaddemane2021}.     

When it comes to phonons, our results generally agree with studies reporting similar lattice parameters and based on GGA-PBE such as \cite{Mounet2018,Gaddemane2021}. The same level of agreement remains when different pseudopotentials are used, with and without spin-orbit coupling. The latter effect is expected to be important for materials with a net magnetic moment \cite{Gaddemane2021}. We do note some small differences around the $\Gamma$-point of the upper optical bands of WS$_2$ when comparing to calculations performed with DFPT \cite{Mounet2018,Gaddemane2021}. 

\begin{figure}[h!] 
  \includegraphics[width=1.0\linewidth]{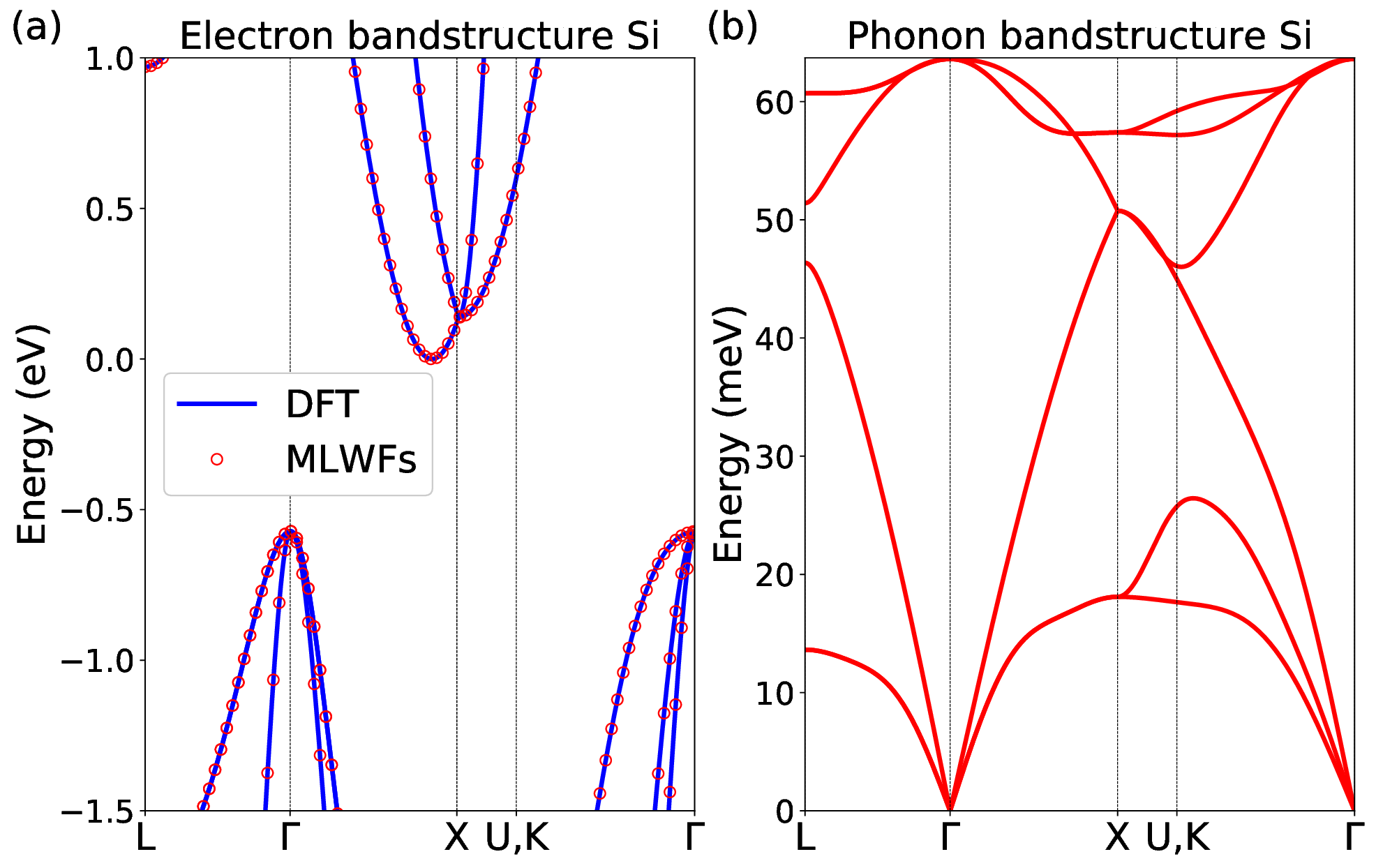}
  \caption{(a) Electronic and (b) phonon band structure of Silicon. The same plotting conventions as in Fig.~\ref{fig:TMDBS} are used.}
  \label{fig:SiBS}
\end{figure}

The calculated electron and phonon band structures of Silicon are depicted in Fig.~\ref{fig:SiBS} and the corresponding band gap given in Table \ref{tab:structureTable}. As in the case of the TMDCs, the calculated band gap (0.57 eV) underestimates its experimental counterpart (1.12 eV) \cite{SzeNg2007}. The Silicon band structure is nevertheless consistent with other theoretical works \cite{Ponce2018,Ma2018,Ponce2020}, and its overall shape in good agreement with experimental measurements \cite{Wei1994}. Improving the band gap accuracy would require applying GW corrections, which is computationally very expensive and not necessary to extract mobility values.            

\subsection{Phonon-Limited Mobility}\label{sec:mob}

\subsubsection{LBTE}

Mobility calculations were carried out with the iterative LBTE method, utilizing homogeneous and commensurate $\textbf{k}$/$\textbf{q}$-point grids across the entire Brillouin Zone. Convergence was tested with respect to the grid density and interaction range, as shown in Appendix A. The results presented here were obtained on a $301 \times 301 \times 1$ $\textbf{k}$/$\textbf{q}$-point grid for the TMDCs and on a $201 \times 201 \times 201$ $\textbf{k}$/$\textbf{q}$-point grid for Silicon. Scattering states up to 350 meV above the conduction band minimum were included, capturing the entire Fermi tail in case of high carrier concentrations. The TMDC mobilities correspond to transport occurring along the zigzag direction. Their values at 300 K are reported in Table \ref{tab:mobTable} for two carrier concentrations, together with the extracted effective masses and the coefficient $\gamma$ representing the temperature dependence of the mobility, i.e., $\mu(T) \propto T^{-\gamma}$ \cite{Kaasbjerg2012}. 

\newcolumntype{C}{>{\centering\arraybackslash}X}
\newcolumntype{D}{>{\centering\arraybackslash}p{0.2\hsize}} 
\begin{table}[h!]
\begin{center}
\begin{tabularx}{\linewidth}{DDDCCC}
\toprule
Material & $\mu_{int}$ & $\mu_{5e13}$ & m$^*$ & m$^*_Q$ & $\gamma$ \\
 & (cm$^2$/(Vs)) & (cm$^2$/(Vs)) & ($m_0$) & ($m_0$) \\
\hline
MoS$_2$  & 221  &  130 & 0.46 & 0.61  & 1.63 \\
MoSe$_2$ & 89   &  53  & 0.53 & 0.55  & 1.54 \\
WS$_2$   & 360  & 186  & 0.31 & 0.60  & 1.27 \\
WSe$_2$  & 260  &  84  & 0.33 & 0.51  & 1.34 \\
Si       & 1452 &  -  & 0.96 &  -     & - \\
\bottomrule 
\end{tabularx}
\end{center}
\caption{Transport properties of the considered materials. The intrinsic mobility ($\mu_{int}$), the mobility at an electron concentration of $5 \cdot 10^{13}$ cm$^{-2}$ ($\mu_{5e13}$), the effective mass at the conduction band minimum (m$^*$), the effective mass along the transport direction at the Q valley (m$^*_Q$), and the coefficient representing the temperature dependence of the mobility ($\gamma$) are reported.}
\label{tab:mobTable}
\end{table}

We report intrinsic phonon-limited mobilities of 221, 89, 360, 260 cm$^2$/Vs for MoS$_2$, MoSe$_2$, WS$_2$, and WSe$_2$, respectively. These results should be put in perspective with the large range of values reported in literature: 47-410 cm$^2$/Vs for MoS$_2$ \cite{Kaasbjerg2012,Li2015,Li2013,Zhao2018,Pilotto2022,Sohier2018,Gaddemane2021,Zhang2014,Huang2016,Jin2014}, 18-269 cm$^2$/Vs for MoSe$_2$ \cite{Gaddemane2021,Zhang2014,Huang2016,Jin2014}, 37-1739 cm$^2$/Vs for WS$_2$ \cite{Sohier2018,Gaddemane2021,Zhang2014,Huang2016,Jin2014}, and 23-1083 cm$^2$/Vs for WSe$_2$ \cite{Sohier2018,Gaddemane2021,Zhang2014,Huang2016,Jin2014}. Although MoS$_2$ has probably been the most studied TMDC, there is till no consensus regarding its true phonon-limited mobility value. The most recent works have narrowed down the range to 120-300 cm$^2$/Vs \cite{Sohier2018,Gaddemane2021,Pilotto2022}. It should however be noted that values in this range correspond to calculations done both at intrinsic and high carrier concentrations. We will discuss the impact of the carrier concentration in the next section.

The $K$-$Q$ valley separation has been identified as a possible source of variability as it influences intervalley scattering. However, attempts to explain the observed discrepancies solely based on this parameter have led to contradictory results. In \cite{Li2013} a decrease in the mobility of MoS$_2$ from 320 cm$^2$/Vs down to 130 cm$^2$/Vs was found upon including intervalley scattering between the $K$ and $Q$ valleys separated by 70 meV, while in \cite{Gaddemane2021} the mobility did not change when varying both the pseudopotentials and the exchange and correlation functional, although the valley separation increased from 100 to 270 meV. This indicates that the $K$-$Q$ valley separation, though important, is only one piece of a more complex interplay of parameters. This feature becomes even more evident when examining WS$_2$, which exhibits the broadest range of calculated mobility values. Part of this variation can be attributed to discrepancies in the effective mass. For example, the study reporting the highest mobility for WS$_2$ (1739 cm$^2$/Vs) relies on an effective mass of 0.26 $m_0$ \cite{Huang2016}. This is nearly 20$\%$ lower than the values both our study and several others have extracted of 0.31 $m_0$ \cite{Gaddemane2021,Sohier2018,Jin2014}. In \cite{Gaddemane2021}, the effective mass at $K$ varies from 0.30 $m_0$ to 0.35 $m_0$, depending on the chosen DFT parameters. They also show that the heavier effective mass of the $Q$ valley can also change between 0.54 $m_0$ and 0.60 $m_0$, which is expected to impact the mobility. Hence, similarly to the $K$-$Q$ valley separation, the effective mass of WS$_2$ shows a high sensitivity to the simulation setup.
 
In summary, sometimes even small changes in the calculated electron and phonon dispersions can have a profound impact on the mobility values. Yet, even with identical dispersions, drastically different mobilities can be obtained. A compelling example is given by \cite{Gaddemane2021}. Specifically, for WS$_2$, they demonstrated that even with comparable band structures and phonon dispersions, a tenfold difference in the electron-phonon matrix-elements can exist if the pseudopotentials differ. While the derived electron energies and phonon frequencies, i.e., their eigenvalues, remain very close, their eigenvectors can be quite different, which has a direct impact on the electron-phonon coupling elements. It can thus be concluded that the selected methodology to determine the scattering rates (here the frozen phonon approach is used as opposed to DFPT) can lead to different results because of the underlying electron and phonon eigenvectors, among other factors. Keeping these facts in mind we believe that our calculated TMDC mobilities agree well with the literature, especially when looking at some of the most recent works for MoS$_2$ \cite{Pilotto2022}, MoSe$_2$ \cite{Gaddemane2021}, WS$_2$ \cite{Gaddemane2021}, and WSe$_2$ \cite{Gaddemane2021}. 

We note that the frozen phonon approach used here does not fully capture long-range interactions such as polar optical phonon scattering (Fr\"ohlich interaction) because it is limited by the supercell size of the DFT calculations. Since the Fr\"ohlich interaction is slow to converge  with respect to the supercell size, it cannot be included \cite{Gunst2016}. However, we have previously demonstrated for MoS$_2$ that including an analytical 2D Fröhlich contribution has a negligible impact on the mobility \cite{Backman2022}. While the Fr\"ohlich contribution may be different in other TMDCs, the agreement between our results and those from studies employing DFPT \cite{Gaddemane2021}, which captures long-range interactions, suggests that it is not a critical factor.

Next, we place our MoS$_2$ simulation finding in the context of experimental measurements. The focus is set on this TMDC because more data is available. The measured mobility of monolayer MoS$_2$ ranges from 23 to 217 cm$^2$/Vs \cite{Yu2016,Cui2015,Liu2015,Yu2014,Sanne2015,Kang2015,Radisavljevic2013,Smithe2018}. In experiments too, multiple parameters can influence the mobility, from the samples themselves to the measurement equipment, the quality of the data, and the dielectric environment. The studies reporting the largest mobility values typically involve high-permittivity gate dielectrics such as HfO$_2$. High-$\kappa$ dielectrics can better suppress Coulomb interactions and, as a consequence, charged impurity scattering (CIS) \cite{Yu2016}. Reducing scattering brings the mobility closer to its intrinsic, phonon-limited value. Our simulations specifically target the mobility of freestanding monolayers and consider electron-phonon scattering, no other mechanisms. Additional scattering processes, notably surface optical phonon (SOPS) and CIS, are known to have a negative impact on the transport properties of TMDCs \cite{Lee2019}. Hence, the calculated MoS$_2$ mobilities accounting for electron-phonon scattering only should be larger than the experimental ones, which suffer from these different scattering sources. With a value of 221 cm$^2$/Vs for monolayer MoS$_2$, we are in a range compatible with experimental data, knowing that the inclusion of SOPS and CIS would decrease the mobility bellow 100 cm$^2$/Vs \cite{Lee2019}. This sanity check gives us confidence in the developed computational framework.

Another possibility to access its accuracy consists of simulating materials for which there are less uncertainties in the mobility value. Silicon emerges as the ideal candidate to do that and to further validate our methodology. We obtain a phonon-limited electron mobility of 1452 cm$^2$/Vs for bulk Silicon. This result aligns very well with with experimental values, which lie between 1300 and 1450 cm$^2$/Vs \cite{Canali1975,Norton1973,SzeNg2007}. On the modeling side, the range is broader and encompasses values between 1080 and 1970 cm$^2$/Vs \cite{Rhyner2013,Fiorentini2016,Li2015,Restrepo2009}. Recent investigations narrowed down this range to 1305-1555 cm$^2$/Vs \cite{Ponce2018}. This study meticulously analyzed the influence of various simulation parameters on the mobility. Different exchange and correlation functionals, spin-orbit coupling, and GW quasiparticle corrections were tested. It was found that the most reliable theoretical estimate for the mobility of bulk Silicon is 1366 cm$^2$/Vs. Interestingly, when employing simulation parameters similar to ours — specifically, the use of GGA, no spin-orbit coupling, and no crystal relaxation (experimental lattice constant) — they obtained a  mobility of 1457 cm$^2$/Vs, a value almost identical to ours. This excellent agreement with both experimental data and the calculations of \cite{Ponce2018} reinforces our confidence in the accuracy of our modeling approach and demonstrates its application potential beyond just TMDCs.       

\begin{figure*}[t]
  \includegraphics[width=1.0\linewidth]{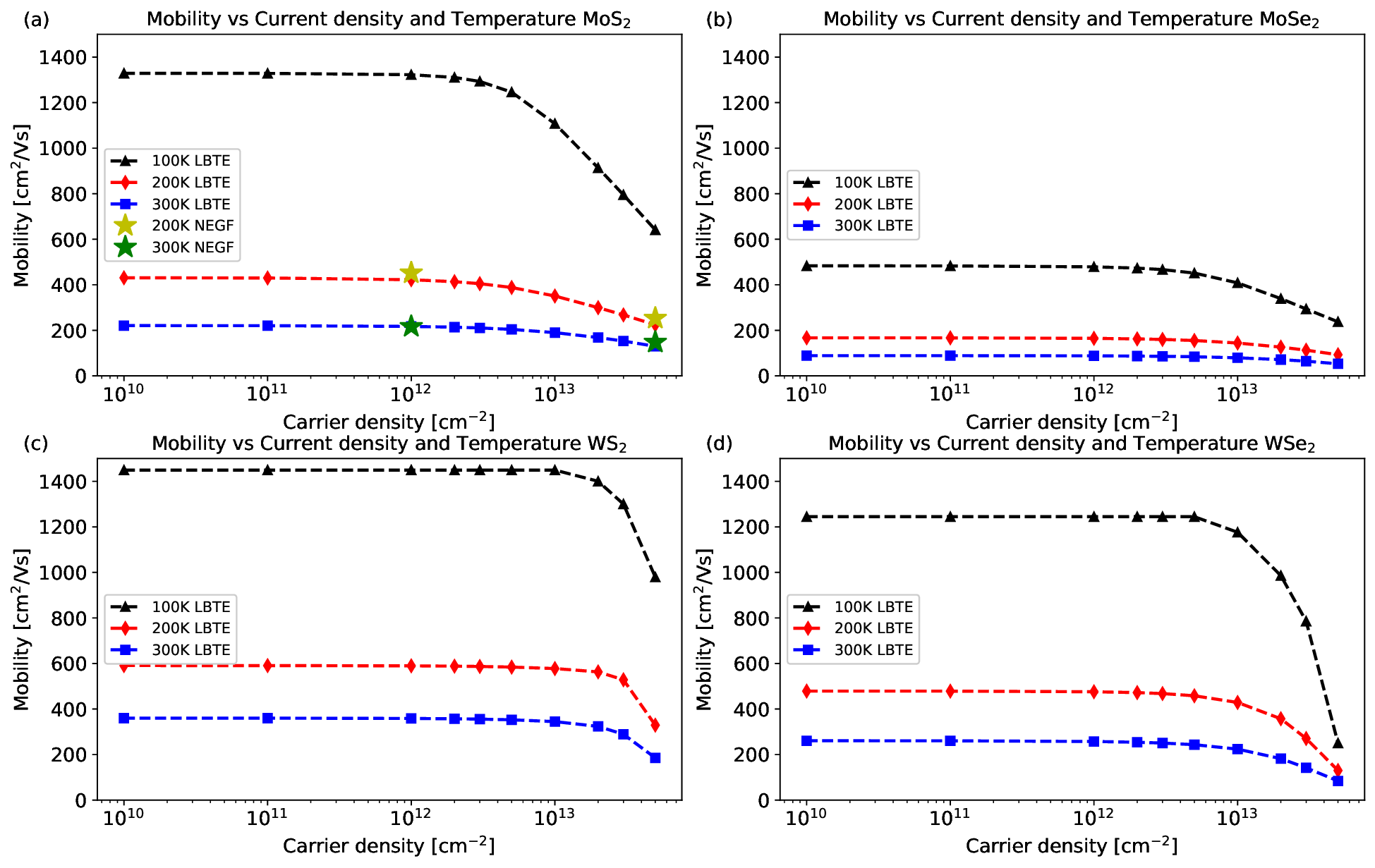}
  \caption{Phonon-limited mobility of the considered TMDCs: (a) MoS$_2$, (b) MoSe$_2$, (c) WS$_2$, and (d) WSe$_2$, as a function of the carrier density for a temperature $T$= 100 K (dashed line with triangles), $T$ = 200 K (dashed line with diamonds), and $T$ = 300 K (dashed line with squares). All these results were obtained with LBTE. The stars in sub-plot (a) refer to NEGF calculations.}
  \label{fig:MobLBTENEGF}
\end{figure*}

So far we have considered mobility calculations under intrinsic carrier concentrations, where the Fermi level is located in the middle of the TMDC band gap. Given the fact that semiconductors are doped in most applications, it is crucial to explore their transport properties at higher carrier concentrations too. To this end, we computed with LBTE the phonon-limited electron mobility of the TMDCs as a function of the carrier density and temperature. Our findings are shown in Fig.~\ref{fig:MobLBTENEGF}. The mobilities were determined as a function of the Fermi level, which was shifted upwards, from the middle of the band gap into the conduction band. The corresponding carrier density was then subsequently computed with Eq.~(\ref{eq:cdLBTE}). The calculated mobilities exhibit the expected behavior: they initially remain flat as the carrier concentration increases up to $10^{12}$ cm$^{-2}$. Above this point higher energy states in the conduction band become populated. The electrons occupying them can relax into low-energy states through optical phonon emission, which increases the scattering rate and decreases the mobility. The observed mobility drop accelerates when the $Q$ valley starts filling up and intervalley scattering kicks in. 

Turning to the temperature dependence of the mobility, we observe that the lower this parameter is, the smaller the phonon population. This leads to a decrease of the scattering rate and to an enhancement of the mobility. In monolayer MoS$_2$, the intrinsic phonon-limited mobility rises from 221 cm$^2$/Vs at 300 K to 431 cm$^2$/Vs at 200 K and further to 1329 cm$^2$/Vs at 100 K. At the same time, the mobility still rapidly drops at high carrier concentrations, in line with what other studies predicted \cite{Kaasbjerg2012,Gunst2016}. Assuming that the mobility evolves as $\mu(T) \propto T^{-\gamma}$ \cite{Kaasbjerg2012}, where $T$ is the temperature, we can extract the parameter, $\gamma$ from our calculations. For MoS$_2$, we find that $\gamma$ = 1.63, a value similar to what other computational studies reported (between 1.35 and 1.69 \cite{Gunst2016,Zhou2021,Kaasbjerg2012}). The calculated $\gamma$ values for the other TMDCs considered in this work are listed in Table \ref{tab:mobTable}.  

For MoS$_2$, we also calculated the intrinsic hole mobility, which came out to be 19 cm$^2$/Vs. Contrary to the electron mobility, fewer simulation studies provide this quantity. For instance, in \cite{Guo2019} a comparable hole mobility of 26 cm$^2$/Vs was obtained, whereas a value of 113 cm$^2$/Vs is reported in \cite{Ponce2023} when incorporating spin-orbit coupling.

\subsubsection{NEGF}
LBTE offers a computationally efficient platform to determine the mobility of bulk and nano-structured materials, but it does not provide information about electrical current or device characteristics. This is where the NEGF formalism can be of great help. Although both approaches take the same quantities as input, they rely on different systems of equations. To validate their implementation, it is therefore useful to compare physical quantities that can be computed with both of them. This is the case of the phonon-limited mobility. Especially, since the scattering self-energies are truncated in NEGF ($\Gamma$-point calculation and introduction of a cut-off radius) while no such approximations are made in LBTE, comparisons between the mobility calculated with both methods allow us to assess the accuracy of the simplifications we made to NEGF.

The phonon-limited mobility is computed in NEGF with the dR/dL method \cite{Rim2002}, which enables consideration of different temperatures and carrier concentrations. We first construct $3\times1\times8$ supercells comprising 144 atoms that are derived from the smallest possible orthorhombic cell made of 6 atoms. These supercells are then replicated along the transport direction $x$ to give rise to devices of lengths roughly equal to 10, 17, and 30 nm. The periodic direction $z$ is modeled via 3 $k_z$-points at low carrier densities (n =$1\cdot10^{12}$ cm$^{-2}$) and 5 $k_z$-points at $n=5\cdot 10^{13}$ cm$^{-2}$. We verified that this number of $k_z$-points is sufficient to capture all band structure features. Such small numbers of $k_z$-points are justified by the large dimension of the transport cell along $z$ ($R_{z}$ = 2.5 nm). The phonon dispersion and eigenvectors are evaluated at the $\Gamma$-point and include a total of 432 phonon energies. In the calculation of the scattering self-energies $\mathbf{\Sigma}_{mn}^{\gtrless ep}$ in Eq.~(\ref{eq:SigEPEnergy}), a cut-off radius $|\textbf{r}_n - \textbf{r}_n| < 12$ \text{\AA} is introduced to keep the computational burden and memory manageable. On average, each atom interacts with 155 neighbors. No larger value could be simulated on the available machine \cite{pizdaint2018}. A detailed analysis of the $r_{cut}$-dependent convergence of the electrical current is provided in Appendix A. All NEGF-based mobility calculations are performed under small bias conditions ($\Delta V = 1$ mV), assuming a flat electrostatic potential. 

\begin{figure}[h]
  \includegraphics[width=1.0\linewidth]{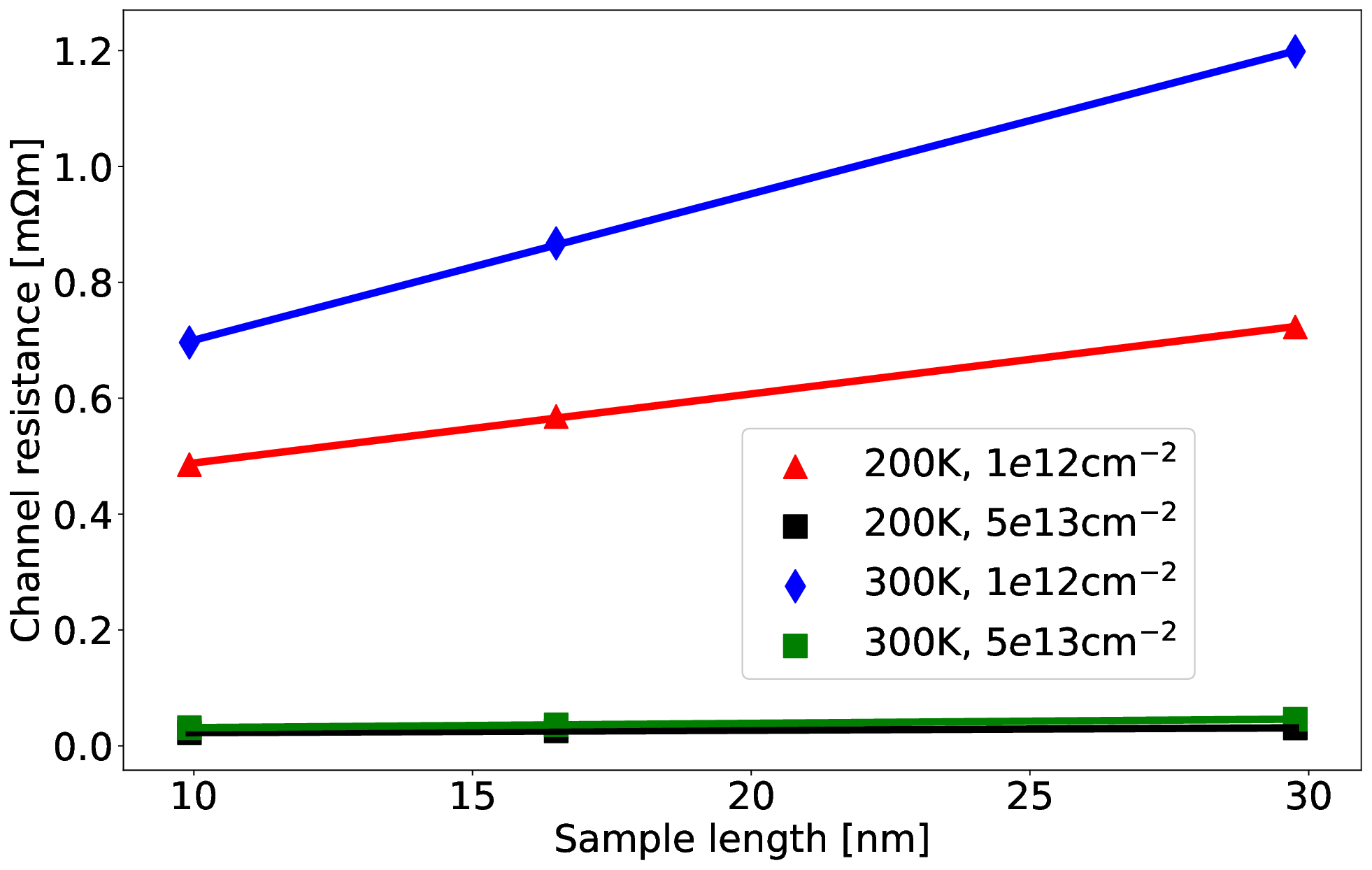}
  \caption{Channel resistance of monolayer MoS$_2$ as a function of the sample length, at temperatures of 200 K and 300 K and for electron concentrations $n = 1\cdot 10^{12}$ cm$^{-2}$ and $n = 5\cdot 10^{13}$ cm$^{-2}$. The symbols represent simulation results, while the lines are used as fits and serve as inputs to the dR/dL method \cite{Rim2002}.}
  \label{fig:RvsL}
\end{figure}
        
Figure \ref{fig:RvsL} presents the obtained MoS$_2$ channel resistances as a function of the device length. Due to the high computational intensity associated with such calculations, we limited ourselves to two temperatures, 200 K and 300 K, and two electron concentrations, $n = 5 \cdot 10^{13}$ cm$^{-2}$ and $n = 1 \cdot 10^{12}$ cm$^{-2}$. Ideally, the sample lengths should be much larger then the mean free path for scattering, but for the same reason as already mentioned above, we could not go beyond 30 nm long devices. Nonetheless, it is obvious from the results in Fig.~\ref{fig:RvsL} that the resistance linearly increases with the sample length in all cases, underscoring the diffusive nature of transport. From this data, the mobility at 300 K and $n = 1 \cdot 10^{12}$ cm$^{-2}$ is found to be 217 cm$^2$/Vs, which is in perfect agreement with our LBTE finding under the same condition (217 cm$^2$/Vs). At a reduced temperature of 200 K, the NEGF mobility rises to 452 cm$^2$/Vs, thus exceeding the LBTE value of 422 cm$^2$/Vs. Meanwhile, increasing the carrier concentration to $n = 5 \cdot 10^{13}$ cm$^{-2}$ reduces the mobility to 148 cm$^2$/Vs at 300 K and 254 cm$^2$/Vs at 200 K in case of NEGF. These results also overestimate the LBTE values, which are equal to 130 cm$^2$/Vs and 224 cm$^2$/Vs at 300 K and 200 K, respectively. The NEGF mobilities are shown in Fig.~\ref{fig:MobLBTENEGF} as stars to enable a direct comparison with the LBTE ones. We believe that the relatively good agreement between LBTE and NEGF (less than 15 \% difference in all cases) validates our approximations to calculate the electron-phonon scattering self-energy. The remaining discrepancies can possibly be explained as follows: considering the lower temperature case first, the higher NEGF mobility can be attributed to the short sample lengths (10, 17, and 30 nm). From the dR/dL results we can estimate the phonon-limited mean free path for carriers \cite{Niquet2014}, which is equal to 9.8 nm at 300 K and 22.1 nm  at 200 K for low carrier concentrations. Given that the mean free path at 200 K is almost on part with the maximum sample length of 30 nm, the channel might not be long enough to let the electron population fully relax due to phonon emission/absorption. An underestimation of the scattering rate is expected to lead to higher mobility values. This reasoning does not apply to the high carrier concentration case, where the mean free path is 6.7 nm at 300 K. In this configuration the difference between LBTE and NEGF may come from the lack of $k_z$-point coupling. While the size of the transport cell ensures that the most relevant states are projected to the $\Gamma$-point, or close to it, this might no more be true at high carrier concentrations. For example, even in the chosen large cell not all $Q$-valley states are projected to $\Gamma$. They however start playing an important role when the electron density reaches $5 \cdot 10^{13}$ cm$^{-2}$. Our $\Gamma$-point calculations neglect parts of the coupling to the Q-valley, which might underestimate the scattering rate and artificially increase the mobility. The error (148 vs. 130 cm$^2$/Vs and 254 vs. 224 cm$^2$/Vs) remains acceptable and indicates that most scattering events are still accounted for.           

\subsection{Device Simulation}
Next, after demonstrating that the NEGF approach can reproduce the LBTE mobilities fairly accurately, we move to full \textit{ab initio} device simulations. The transfer characteristics of a single-gate monolayer MoS$_2$ FET, as depicted in Fig.~\ref{eq:deviceBlocks}, are investigated assuming perfectly ohmic contacts. The gate length, $L_g$ = 11.8 nm, was chosen according to the latest International Roadmap for Devices and Systems (IRDS) for the year 2028 \cite{ieee_irds}. The source and drain extensions measure $L_s$ = $L_d$ = 9 nm and are doped with a donor concentration $N_D$ = $5 \cdot 10^{13}$ cm$^{-2}$. The transistor structure is therefore composed of 2592 atoms. The MoS$_2$ channel is deposited on a SiO$_2$ “substrate” with a top HfO$_2$ dielectric layer of thickness t$_{OX}$ = 3 nm and a relative permittivity $\epsilon_{OX}$ = 20, resulting in an equivalent oxide thickness (EOT) of 0.58 nm. The OFF-state current $I_{\text{OFF}}$ is set to 0.1 $\mu A/\mu m$ by adjusting the gate work function. The same simulation settings as for the mobility calculations at $n = 5 \cdot 10^{13}$ cm$^{-2}$ and  $T$ = 300 K are used (5 $k_z$-points, an energy resolution of 2 meV).

\begin{figure}[h]
  \includegraphics[width=1.0\linewidth]{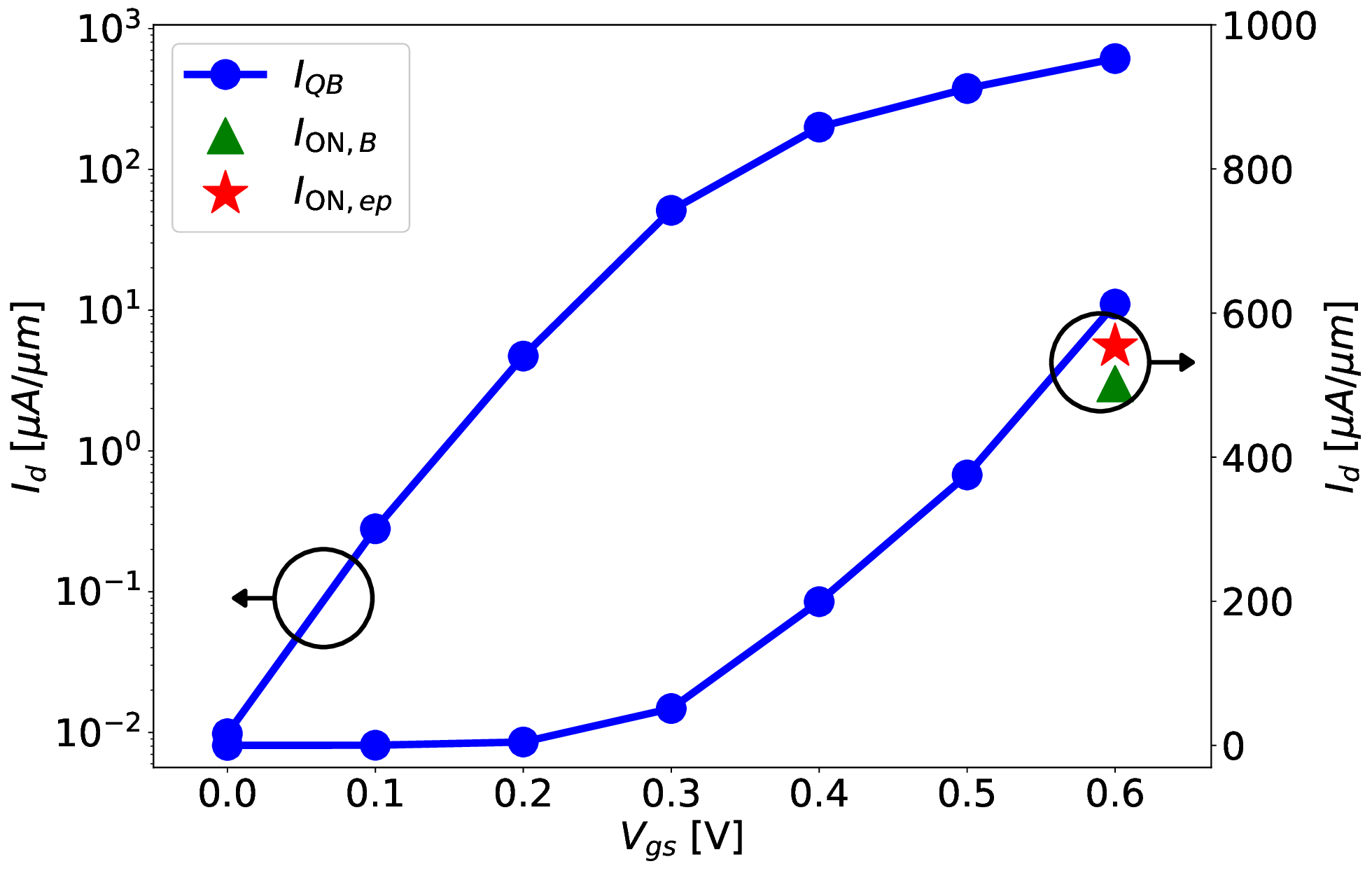}
  \caption{Transfer characteristics $I_{d}$-$V_{gs}$ at $V_{ds}$ = 0.6 V of a single-gate monolayer MoS$_2$ field-effect transistor in the quasi-ballistic limit of transport ($I_{QB}$, blue curve with circles). The ON-state current in the pure ballistic limit ($I_{\text{ON},B}$, green triangle) and in the presence of electron-phonon interactions according to Eq.~(\ref{eq:SigEPEnergy}) ($I_{\text{ON},ep}$, red star) are indicated. Both a linear and a log scale are provided for $I_{QB}$.} 
  \label{fig:IVMoS2}
\end{figure}

First, the $I_{d}$-$V_{gs}$ transfer characteristics of the MoS$_2$ transistors are simulated in the quasi-ballistic limit of transport. The phenomenological electron-phonon scattering model of \cite{Klinkert2020} is recalled for that purpose with a phonon frequency $\hbar \omega$ = 40 meV and a scattering strength $D_{ep}$ = 25 (meV)$^2$. This model ensures that the non-physical negative differential behavior often observed in TMDC-based FET is eliminated. The resulting curve at $V_{ds}$ = 0.6 V is depicted in Fig.~\ref{fig:IVMoS2}. It can be seen that the device exhibits good electrostatic properties with a sub-threshold swing of 69 mV/dec and a decent ON-state current ($I_{d}$ at $V_{ds}$ = $V_{gs}$ = 0.6 V) of 613 $\mu A/\mu m$. It is also found that the considered transistor does not suffer from the so-called density-of-states bottleneck \cite{Fischetti2007}, the extracted gate capacitance $C_{g}= \partial n/ V_{gs} =  5.6$ $\mu F/cm^2$ approaching the value of the oxide capacitance ($C_{ox}=5.9$ $\mu F/cm^2$).

As computing the entire $I_{d}$-$V_{gs}$ in the presence of electron-phonon scattering as described in Eq.~(\ref{eq:SigEPEnergy}), is computationally not feasible, we focus on the ON-state and evaluate it with the same electrostatic potential as in the case of the phenomenological model. As a reference, we also determined the value of the pure ballistic current by turning off all scattering interactions. Both the pure ballistic and the dissipative currents are displayed in Fig.~\ref{fig:IVMoS2} as a triangle (pure ballistic current, $I_{\text{ON},B}$ = 502 $\mu A/\mu m$) and as a star (current with real electron-phonon scattering, $I_{\text{ON},ep}$ = 554 $\mu A/\mu m$). It can be noticed that the $I_{\text{ON},B}$ is smaller than the ON-state currents obtained in the presence of scattering. This phenomenon was explained in \cite{Szabo2015EP}: phonons connect electronic bands with a narrow energy width that cannot carry current in the ballistic limit of transport. In this case, electron-phonon scattering can increase the current. There is however a competing effect that manifests itself at high scattering rates: back-scattering or the decrease of the current caused by electron-phonon interactions. This is exactly what happens when the full electron-phonon scattering model is turned on: back-scattering starts dominating so that the current decreases as compared to the quasi-ballistic case. It nevertheless remains higher than in the pure ballistic case. Overall, the difference in current is rather small, $I_{\text{ON},B}$ = 502 $\mu A/\mu m$ and $I_{\text{ON},ep}$ = 554 $\mu A/\mu m$, but the electron behavior significantly differs. 

\begin{figure}[h!]
  \includegraphics[width=1.0\linewidth]{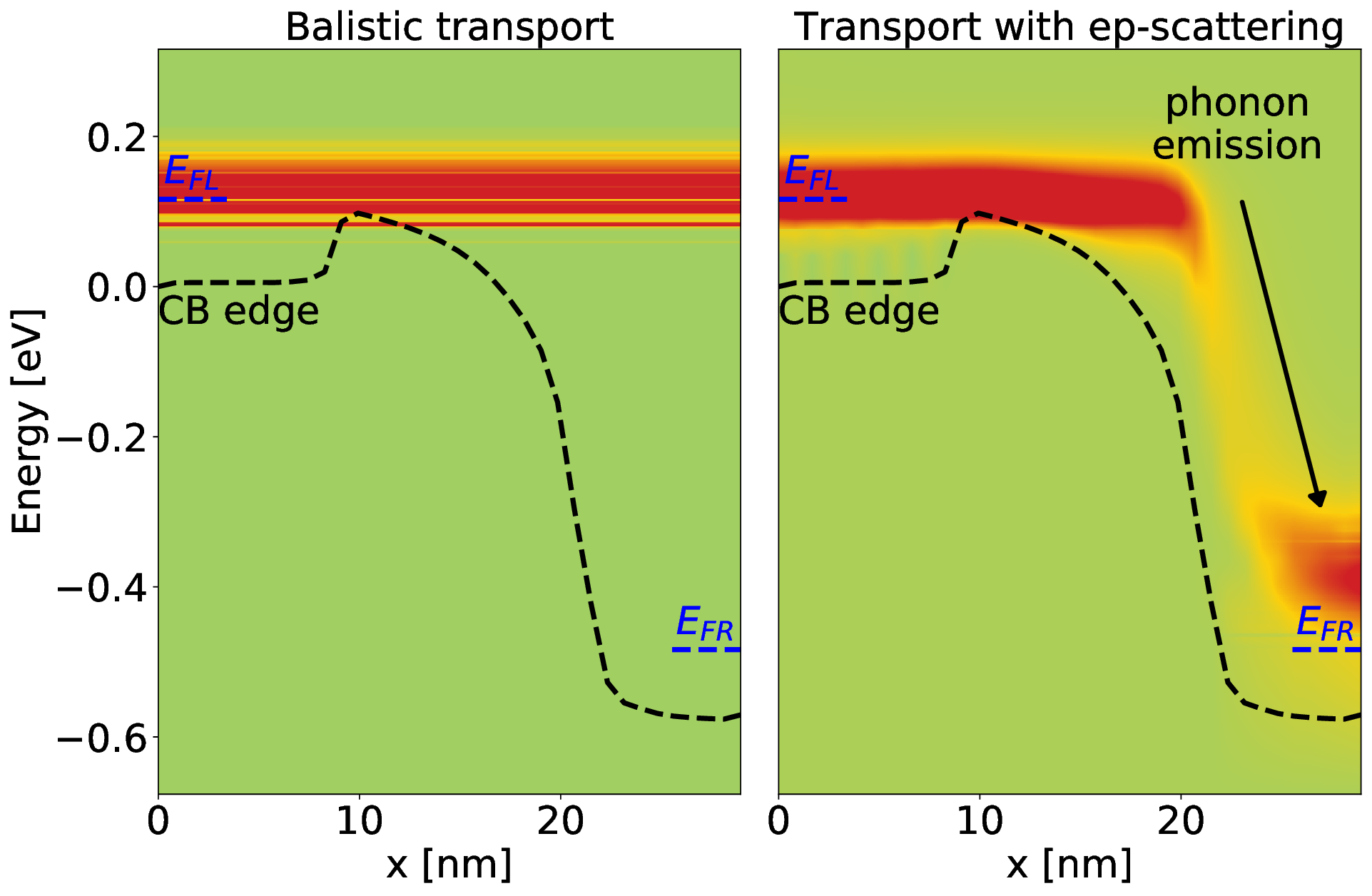}
  \caption{Spectral ON-state current ($I_{d}$ as a function of energy $E$ and position $x$) for a single-gate monolayer MoS$_2$ field-effect transistor in the ballistic limit of transport (left) and when the proposed electron-phonon scattering model is turned on (right). Red indicates high current concentration, while no current flows through the green regions. The conduction band edge is marked by a black dashed line. The Fermi levels of the left ($E_{\text{FL}}$) and right ($E_{\text{FR}}$) contacts are highlighted in blue.}    
  \label{fig:IvsEMoS2}
\end{figure}
               
This is exemplified in Fig. \ref{fig:IvsEMoS2} where the energy- and position-resolved I$_{\text{ON},B}$ and I$_{\text{ON},ep}$ are plotted. In the ballistic case the spectral current remains constant from the left side (source) to the right side (drain): the energy of the electrons injected into the simulation domain cannot vary in the absence of a dissipative scattering mechanism. When electron-phonon interactions are introduced, electrons lose a substantial amount of their energy between the source and the drain of the transistor. Most of the energy relaxation occurs in the region located after the top of the potential barrier, where the electric field reaches its maximum. 

The effect of back-scattering, which caused electrons flowing from the source to the drain to be scattered back to their origin, cannot be directly visualized, but its influence can be indirectly measured when considering the injection velocity extracted at the top of the barrier. This velocity decreases from $3.6 \cdot 10^{6}$ cm/s in the quasi-ballistic case to $3.2 \cdot 10^{6}$ cm/s when electron-phonon scattering is fully accounted for. Note that in ultra-scaled devices, the injection velocity and charge at the top of the potential barrier are not as clearly defined as in their long-channel counterparts. This comes from the absence of a plateau in the channel's center. The extracted injection velocity values tend to vary depending on the interpolation technique that is used to identify the exact location of the top of the potential barrier. 
       
\section{Conclusion and Outlook}\label{sec:Con}

This work explored the transport properties of selected TMDCs and bulk Silicon  based on an \textit{ab initio} description of the their electron and phonon dispersions and of the interaction between these particles. The developed methodology relies on a combination of first-principles DFT, MLWF, LBTE, and NEGF. Mobility calculations reveal that the obtained results align very well with the current literature, despite the myriad of factors that may impact the calculated quantities. The sensitivity of the TMDC mobility to factors such as pseudopotentials, exchange-correlation functionals, and phonon calculation method was highlighted. Hence, as a sanity check, the mobility of Silicon, a well characterized semiconductor, was computed and compared to experimental data. The excellent agreement between our calculations, measurements, and other computational studies validates our treatment of electron-phonon scattering. The latter was therefore employed in NEGF to construct electron-phonon scattering self-energies going beyond the diagonal approximation, but restricted to $\Gamma$-point calculations. Here again, good agreement between LBTE and NEGF indicates that the most important features of electron-phonon scattering are captured by our NEGF solver. As a consequence, \textit{ab initio} device simulations of TMDC-based FETs incorporating the real mobility of these materials could be performed. Being able to precisely account for electron-phonon scattering in the investigation of TMDC devices is essential to shed light on their intrinsic transport properties. While the the accuracy achieved by our NEGF approach for mobility calculations and device simulations is promising, further enhancements are required to make it readily applicable to any system. Its computational efficiency should be improved so that it can treat larger devices and structures embedded within a dielectric environment \cite{Fiore2022}. Such an extension is key to directly account for surface optical phonon scattering. Significant speedups might be possible by applying the mode-space approximation \cite{Ducry2020} and by projecting the electron-phonon scattering self-energies into this basis. Additionally, the possibility to go beyond the diagonal approximation for scattering self-energies can now be used to treat other mechanisms as well, for example (charged) impurity scattering, surface roughens, or alloy disorder. The inclusion of non-diagonal scattering entries ensures a more detailed and accurate description of carrier interactions with their environment.

\section{Acknowledgments}

This research was supported by NCCR MARVEL, funded by the Swiss National Science Foundation (SNSF) under grant No. 182892, by grant No. 175479 for SNSF (ABIME), and by the Swiss National Supercomputing Center (CSCS) under projects s1119 and s1212.

\section{Declaration of competing interest}

The authors declare that they have no known competing financial interests or personal relationships that could have appeared to influence the work reported in this paper.

\appendix
\section*{Appendix A: Convergence analysis}\label{sec:AppenA}

To ensure the accuracy of the electrical results presented in this paper, we evaluate here the convergence of the calculated mobility and electrical currents, using monolayer MoS$_2$ as a representative example, with respect to the to $\textbf{k}$/$\textbf{q}$-point grid (LBTE) and interaction range (LBTE and NEGF). 

\begin{figure}[h!]
  \includegraphics[width=1.0\linewidth]{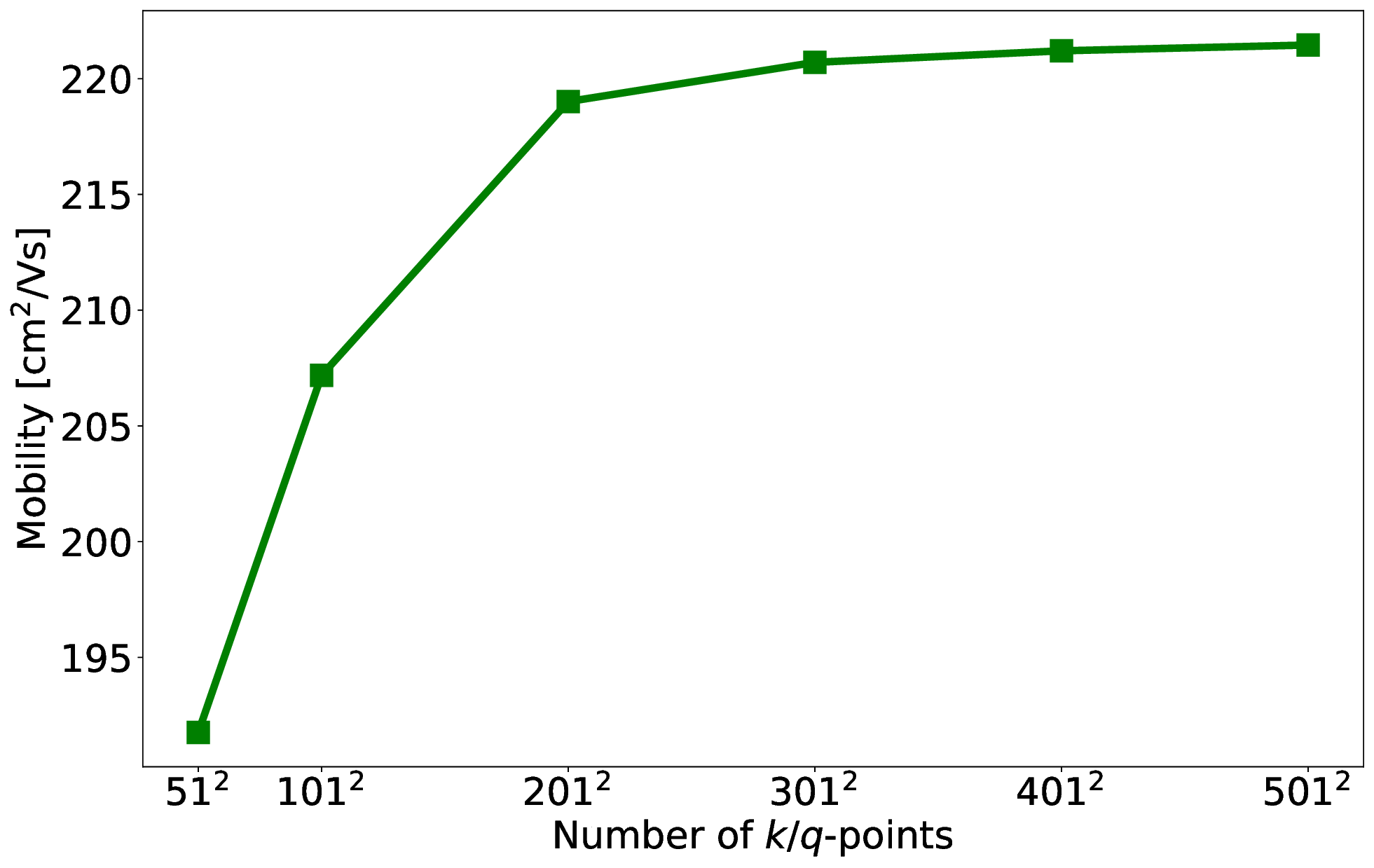}
  \caption{Phonon-limited mobility of monolayer MoS$_2$ as a function of the $\textbf{k}$/$\textbf{q}$-point density. At a $301 \times 301 \times 1$ $\textbf{k}$/$\textbf{q}$-point grid, the phonon-limited mobility shows only a 0.3\% deviation from the value obtained with a denser $501 \times 501 \times 1$ $\textbf{k}$/$\textbf{q}$-point grid.}     
  \label{fig:Mobvsk}
\end{figure}

The implemented LBTE solver enables us to sample the Brillouin zone with high resolution, extending up to $501 \times 501 \times 1$ $\textbf{k}$/$\textbf{q}$-point grids for TMDCs. The convergence of the phonon-limited mobility of MoS$_2$ is plotted in Fig.~\ref{fig:Mobvsk} as a function of the $\textbf{k}$/$\textbf{q}$-point density used in the LBTE calculations. It can be seen that a $301 \times 301 \times 1$ $\textbf{k}$/$\textbf{q}$-point grid is sufficient to yield converged results, the mobility at this grid size only displaying a 0.3\% deviation from the densest grid.

\begin{figure}[h!]
  \includegraphics[width=1.0\linewidth]{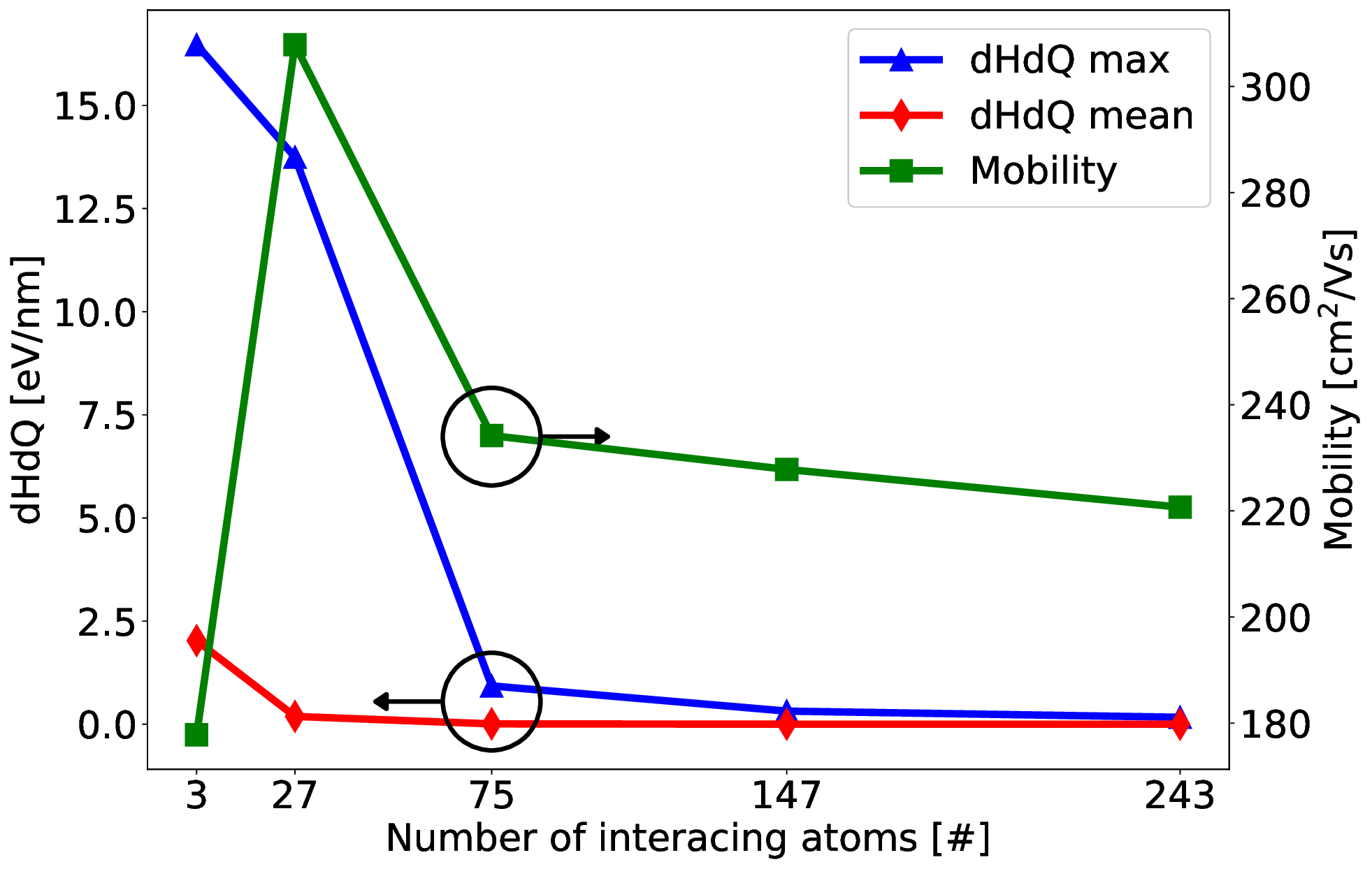}
  \caption{Phonon-limited mobility of monolayer MoS$_2$ as calculated with LBTE (green line with squares), the maximum value of the Hamiltonian derivative dH/dQ (blue line with triangles), and the mean value of dH/dQ (red line diamonds) are reported as a function of the number of atoms interacting with an arbitrary point of reference.}    
  \label{fig:Mobvsrcut}
\end{figure}

In the case of LBTE, the interaction range used to construct the electron-phonon coupling elements can be effectively reduced by constraining the summation over neighboring cells, ($\beta,\gamma$) in Eq.~(\ref{eq:M1}). In Fig.~\ref{fig:Mobvsrcut} we show the resulting mobility as a function of the number of interacting atoms included. It can be observed that the mobility, with interactions limited to 75 atoms, already approaches its converged value with approximately 6\% difference as compared to the case where the complete data set is incorporated. This discrepancy diminishes to below 3\% when interactions with up to 147 atoms are included. Relatively rapid convergence is also observed in the localized Hamiltonian derivatives, dH/dQ$_{I\eta}$ in Eq.~(\ref{eq:M1}). Their maximum and mean values are reported in Fig.~\ref{fig:Mobvsrcut} as a function of the number of interacting atoms. As expected, these derivatives quickly decrease and remain small when more than 75 atoms are included. When the full data is considered the magnitude of the additional long-range interactions becomes comparable to the noise-floor of the MLWF calculations.      

\begin{figure}[h!]
  \includegraphics[width=1.0\linewidth]{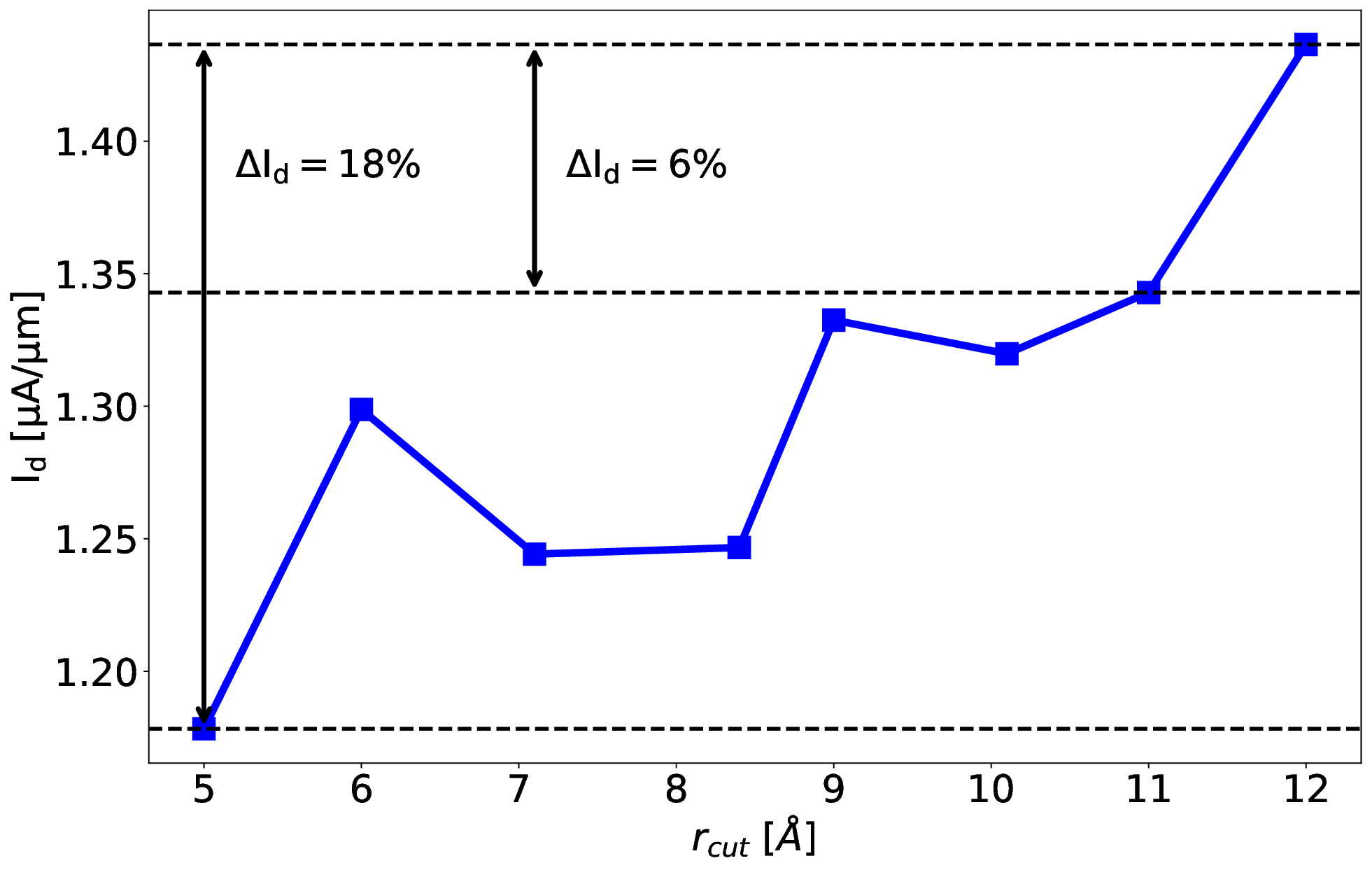}
      \caption{Electrical current flowing through a 10 nm long monolayer MoS$_2$ sample under flat band potential and an applied voltage of 1 mV as a function of the cut-off radius $r_{cut}$ applied to Eq.~(\ref{eq:SigEPEnergy}). The relative changes of the current between $r_{cut}$ = 5 and 12 \text{\AA} (18\%) and between $r_{cut}$ = 11 and 12 \text{\AA} (6\%) are indicated.}    
  \label{fig:Idvsrcut}
\end{figure}

Moving on to the NEGF calculations, we compared the electrical current flowing through a 10 nm long MoS$_2$ sample with 3 $k_z$-points as a function of the interaction range $r_{cut}$. Our interaction range refers to maximum distance between two interacting atomic orbitals. Beyond $r_{cut}$, the scattering self-energies are assumed to be zero. It should however be noted that our NEGF simulations incorporates all available atomic displacements ($I \eta$) in Eq.~(\ref{eq:M1}), regardless of $r_{cut}$. The convergence behavior is reported in Fig.~\ref{fig:Idvsrcut} for $r_{cut}$ between 5 and 12 \text{\AA}, the maximum value that could be reached on the available hardware. When $r_{cut}$ increases from 5 to 12 \text{\AA}, the current changes by 18 \% only, without converging towards a fixed value. At this point, we have no explanation for this behavior. As the mobilities calculated with  $r_{cut}$ = 12 \text{\AA} agree fairly well with those of LBTE, we can only assume that a sufficiently large number of non-diagonal entries in the scattering self-energies are taken into account. Further investigations will be conducted after getting access to a larger machine or after projecting the Hamiltonian and electron-phonon coupling elements onto a mode space basis. 

\bibliography{EP.bib}


\end{document}